\documentclass[%
reprint,
 amsmath,amssymb,
 aps,
prb,
]{revtex4-2}

\usepackage{color}
\usepackage{graphicx}
\usepackage{dcolumn}
\usepackage{bm}
\usepackage{braket}

\begin{document}

\preprint{APS/123-QED}

\title{
Nonreciprocal magnon excitations by the Dzyaloshinskii-Moriya interaction\\ on the basis of bond magnetic toroidal multipoles
}

\author{Takuya Matsumoto}
\author{Satoru Hayami}%
 \affiliation{%
 Department of Applied Physics, The University of Tokyo, Tokyo 113-8656, Japan
 }%

\date{\today}
\begin{abstract}
The microscopic conditions for the emergence of nonreciprocal magnon excitations in noncentrosymmetric magnets are theoretically investigated.
We show that asymmetric magnon excitations appear when a bond magnetic toroidal dipole becomes active, which is defined as a parallel component between the Dzyaloshinskii-Moriya vector and the averaged spin moments at the ends of the bonds. 
Depending on magnetic structures accompanying the bond magnetic toroidal dipoles, the higher-rank magnetic toroidal multipoles can also be activated in a magnetic cluster, which describes various angle dependences of asymmetric magnon excitations. 
We demonstrate a variety of asymmetric magnon excitations for two magnetic systems in the noncentrosymmetric lattice structures; a one-dimensional breathing chain under a uniform crystalline electric field and a three-dimensional layered breathing kagome structure. 
We show that a bottom shift of magnon bands occurs by the magnetic toroidal dipole and a valley splitting occurs by the magnetic toroidal octupole under magnetic orderings.  
\end{abstract}

\maketitle


\section{introduction}
Spatial inversion symmetry is one of the most fundamental symmetries in condensed matter physics. 
Once inversion symmetry is broken in the lattice structure or by electronic orderings through electron correlation, various fascinating phenomena are induced, such as the spin Hall effect~\cite{sinova2015spin}, multiferroics~\cite{fiebig2016evolution}, and noncentrosymmetric superconductivity~\cite{bauer2004heavy}. 
The key ingredient in these phenomena is an antisymmetric spin-orbit interaction that arises from the relativistic spin-orbit coupling in the absence of inversion symmetry.
The antisymmetric spin-orbit interaction results in the modulation of the electronic band structures so as to have a spin polarization at each wave vector in an antisymmetric way, which is termed the spin-momentum locking;
there is a coupling as $\mathbf{k} \times \bm{\sigma}$ ($\mathbf{k}$ is a wave vector and $\bm{\sigma}$ is an electron spin) in polar magnets~\cite{rashba1960properties}, while there is a coupling as  $\mathbf{k}\cdot \bm{\sigma}$ in chiral magnets~\cite{PhysRevLett.115.026401}.

Similar band modulations also occur in other quasiparticle systems.
For example, a magnon, which is a quasiparticle in magnetic systems, exhibits an asymmetric (nonreciprocal) band modulation in the absence of spatial inversion symmetry~\cite{PhysRevLett.30.125, doi:10.1143/JPSJ.56.3635, doi:10.7566/JPSJ.85.053705, PhysRevB.92.184419, zhang2015plane, cho2015thickness, PhysRevB.94.144420, PhysRevB.93.235131, PhysRevLett.119.047201, PhysRevB.95.220406,  tacchi2017interfacial, chaurasiya2018dependence, PhysRevB.97.224403,PhysRevB.98.064416, sato2019nonreciprocal,PhysRevB.101.224419}.
The key ingredient for the nonreciprocal magnon band dispersions is the Dzyaloshinskii-Moriya (DM) interaction~\cite{DZYALOSHINSKY1958241, PhysRev.120.91}. 
This is a counterpart of the antisymmetric spin-orbit interaction in the electron systems, which becomes a source of various phenomena in magnetic insulating systems, such as the magnon Hall effect~\cite{onose2010observation} and the optical magnetoelectric effect~\cite{saito2008gigantic}.
The nonreciprocal magnon dispersions have been observed in experiments~\cite{PhysRevB.92.184419, zhang2015plane, cho2015thickness, PhysRevB.94.144420, PhysRevB.93.235131, PhysRevLett.119.047201, PhysRevB.95.220406,  tacchi2017interfacial, chaurasiya2018dependence, PhysRevB.98.064416, sato2019nonreciprocal}, e.g., for the noncentrosymmetric ferromagnet $\mathrm{LiFe_5O_8}$~\cite{PhysRevB.92.184419} and the antiferromagnet $\alpha$-$\rm{Cu_2 V_2O_7}$~\cite{PhysRevB.95.245119, PhysRevLett.119.047201, PhysRevB.96.180414}. 
Furthermore, physical phenomena caused by the nonreciprocal magnons have been theoretically investigated, such as the nonreciprocal optical response~\cite{PhysRevLett.114.197203, PhysRevB.89.195145, Miyahara2013, doi:10.1143/JPSJ.81.023712, takahashi.Nat.Phys., PhysRevB.99.094401} and the nonreciprocal spin Seebeck effect~\cite{PhysRevB.98.020401, PhysRevB.96.180414}.

Motivated by these studies, we here examine a microscopic origin of the nonreciprocal magnons in more details by focusing on the role of the DM interaction.
For that purpose, we introduce a concept of magnetic toroidal multipoles, which can be an indicator of the nonreciprocal magnons induced by the DM interaction~\cite{doi:10.7566/JPSJ.85.053705}. 
We show that the emergence of the nonreciprocal magnons corresponds to an active bond magnetic toroidal dipole (BMTD) in a model Hamiltonian. 
The BMTD is defined for each bond as a parallel component between the DM vector and averaged spin moments at the ends of the bonds. 
By evaluating such a quantity for all the bonds, one can obtain a spatial distribution of the BMTD in a magnetic cluster referred to as a cluster magnetic toroidal (CMT) multipoles, which determines a momentum-space functional form of nonreciprocal magnons.
Our scheme can be applied to noncentrosymmetric systems with the DM interaction irrespective of collinear and noncollinear magnetic orderings and spatial dimensions. 
To demonstrate that, we analyze two systems. 
One is a collinear magnetic ordered state on a one-dimensional spin chain under a uniform crystalline electric field. 
The other is collinear and noncollinear magnetic ordered states in a three-dimensional breathing kagome structure. 
We show that different types of nonreciprocal magnons are found in different magnetic orderings, which are classified by the type of the CMT multipoles.

The organization of this paper is as follows.
In Sec.~\ref{oneD}, we analyze the one-dimensional spin chain and introduce the BMTD. 
In Sec.~\ref{kagome},  we apply the BMTD to the three-dimensional breathing kagome system. 
Section~\ref{summary} is devoted to a summary of the present paper.
In Appendices~\ref{1DGS} and \ref{GS2}, we discuss the classical spin configurations in the one-dimensional spin chain and the three-dimensional breathing kagome system, respectively.
In  Appendix~\ref{SW2}, we show the Bogoliubov Hamiltonian in the three-dimensional  breathing kagome system.
In Appendix~\ref{other}, we present the nonreciprocal magnon dispersions under the magnetic orderings in the three-dimensional breathing kagome system, which are not discussed in the main text. 

\section{one-dimensional spin chain}
\label{oneD}
In this section, we introduce a concept of BMTD, which gives a quantity to investigate the nonreciprocal magnon excitations, by analyzing a one-dimensional spin chain.
We present a localized spin model with the staggered exchange interactions in Sec.~\ref{model1} and outline a linear spin-wave calculation on the basis of the Holstein-Primakoff transformation in Sec.~\ref{SW1}. 
In Sec.~\ref{result1}, we show that the nonreciprocal magnon excitations are related to the appearance of the BMTD in both the ferromagnetic and antiferromagnetic states.

\begin{figure}[h]
\begin{center}
\includegraphics[width=1.00\linewidth]{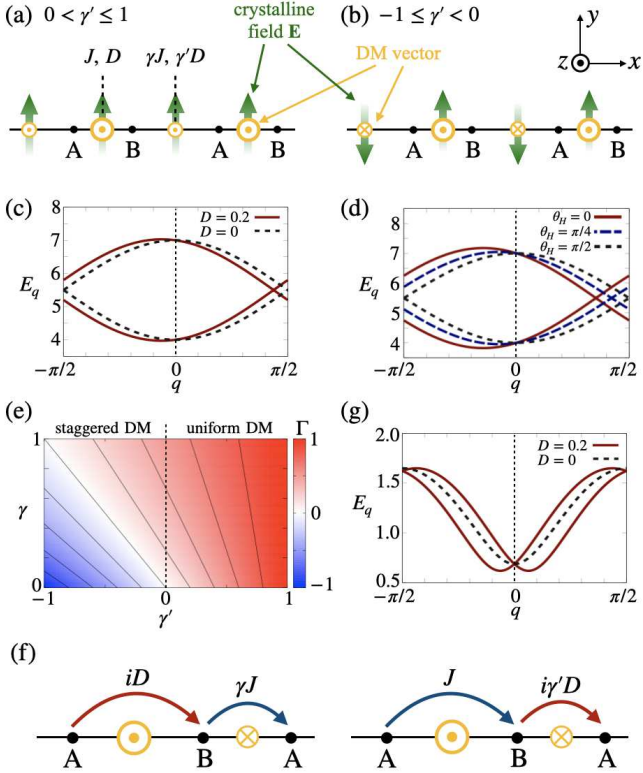}
\end{center}
\caption{
Schematic pictures of one-dimensional spin chains with (a) uniform and (b) staggered DM interactions perpendicular to the crystalline field $\mathbf{E}$.
(c,d) Nonreciprocal magnon excitations in the ferromagnetic order at (c) $\theta_{H}=0$ for $D=0$ and 0.2 and (d) $D=0.5$ for $\theta_{H}=0$, $\pi/4$, and $\pi/2$.
The other parameters are $(J,H,K,\gamma, \gamma')=(-1,4,0,0.5,0.5)$.
(e) $\gamma$ and $\gamma'$ dependences of $\Gamma =(\gamma+\gamma')/(1+\gamma)$.
Contour lines are drawn every $0.2$.
(f) The magnon hopping processes relevant to the nonreciprocal dispersions for $\gamma'<0$.  
(g) Nonreciprocal magnon excitations in the antiferromagnetic order at $(J, H, \theta_{H}, K, \gamma, \gamma')=(1, 0, 0, 0.1, 0.5, 0.5)$ for $D=0$ and $0.2$. 
}
\vspace{5mm}
\label{fig1} 
\end{figure}

\subsection{Spin model}
\label{model1}

Let us start by considering a localized spin model in the one-dimensional system consisting of the alternating bonds as shown in Fig.~\ref{fig1}(a). 
 We take the bond length of both A-B and B-A bonds unity for simplicity, i.e. the norm of primitive transition vector is 2.
The difference of the two bonds is taken as the different magnitude of the interactions.
The model Hamiltonian is given by
\begin{align}
\label{eq1}
\mathcal{H}=&\sum_i 
\Big[J\mathbf{S}_{i \rm{A}}\cdot \mathbf{S}_{i \rm{B}}
+K S^z_{i \rm{A}}S^z_{i \rm{B}}
+D(\mathbf{S}_{i\rm{A}}\times \mathbf{S}_{i\rm{B}})^z\nonumber\\
&+\gamma J\mathbf{S}_{i \rm{B}} \cdot \mathbf{S}_{i+1 \rm{A}}
+\gamma K S^z_{i \rm{B}}S^z_{i+1 \rm{A}}
+\gamma' D(\mathbf{S}_{i\rm{B}}\times \mathbf{S}_{i+1\rm{A}})^z\nonumber\\
&-\mathbf{H}\cdot(\mathbf{S}_{i\rm{A}}+\mathbf{S}_{i\rm{B}})\Big],
\end{align}
where $\mathbf{S}_{i\eta}$ is the classical spin at site $i$ for sublattice $\eta=$A and B.   
By denoting $(i, \eta)$ for the spin with the sublattice $\eta$ and the $i$th unit cell,
the first (second) and fourth (fifth) terms represent the isotropic (anisotropic Ising-type) exchange interactions for the bonds between $(i, {\rm A})$ and $(i, {\rm B})$ spins and $(i+1, {\rm A})$ and $(i, {\rm B})$ spins, respectively.
The coupling constants are given by $J$ and $K$ for the bonds between $(i, {\rm A})$ and $(i, {\rm B})$, while they are given by $\gamma J$ and $\gamma K$ for the bonds between $(i+1, {\rm A})$ and $(i, {\rm B})$, where $\gamma$ stands for a breathing parameter for the alternating bonds.
The third and sixth terms represent the antisymmetric anisotropic exchange interactions, which correspond to the DM interaction. 
We here assume that the DM interaction originates from a polar crystalline electric field or an external electric field along the $y$ direction so that the DM vector lies along the $z$ direction.
The coupling constants are given by $D$ for the bonds between $(i, {\rm A})$ and $(i, {\rm B})$ and $\gamma' D$ for those between $(i+1, {\rm A})$ and $(i, {\rm B})$. 
The last term is the Zeeman interaction under an external magnetic field in the $xz$ plane with the magnitude $H$, $\mathbf{H}=(H_x,0,H_z)=H(\sin\theta_{H},0,\cos\theta_{H})$. 

In the following calculations, we consider two situations: one is the positive $\gamma'$ ($0<\gamma' \leq 1$) and the other is the negative $\gamma'$ ($-1 \leq \gamma' < 0$), while $\gamma$ is always taken to be positive. 
In the former case, the DM vectors on the alternating bonds align the same direction, as shown in Fig.~\ref{fig1}(a), whereas they align in the opposite directions in the latter case, as shown in Fig.~\ref{fig1}(b).

\subsection{Bogoliubov Hamiltonian}
\label{SW1}
We consider the magnon excitations of the model in Eq.~(\ref{eq1}) by supposing that the ground-state spin configuration is described by the two-sublattice ordering as $\mathbf{S}_{i\eta}=(\sin{\theta_{\eta}},0,\cos{\theta_{\eta}})$.
We apply the Holstein-Primakov transformation within a liner spin wave approximation, which is given by $\tilde{S}_{i\eta}^{+}=\sqrt{2S}\eta_i$, $\tilde{S}_{i\eta}^{-}= \sqrt{2S}\eta_i^{\dagger}$, and $\tilde{S}_{i\eta}^z=S-\eta_i^{\dagger}\eta_i$.  Here, $\eta_i = a_i$ and $b_i$ are boson operators for sublattices A and B, respectively, and $\tilde{\mathbf{S}}_{i\eta}$ is directed to the local quantization axis as
\begin{align}
\label{eq2}
\begin{pmatrix}
S^{x}_{i\eta}\\
S^{y}_{i\eta}\\
S^{z}_{i\eta}
\end{pmatrix}
=
R_y (\theta_{\eta})
\begin{pmatrix}
\tilde{S}^{x}_{i\eta}
\\
\tilde{S}^{y}_{i\eta}
\\
\tilde{S}^{z}_{i\eta}
\end{pmatrix},
\end{align}
where $R_y(\theta)$ is the rotation matrix around the $y$ axis by $\theta$.
By substituting Eq.~(\ref{eq2}) into Eq.~(\ref{eq1}) and performing the Holstein-Primakov transformation, we obtain the Bogoliubov Hamiltonian as
\begin{align}
\label{eq3}
H=\frac{S}{2}\sum_{q}\Psi^{\dagger}_{q}
\begin{pmatrix}
Z_{\rm{A}}&F_{q}&0&G_{q}\\
F^{*}_{q}&Z_{\rm{B}}&G_{-q}&0\\
0&G^{*}_{-q}& Z_{\rm{A}}&F^{*}_{-q}\\
G^{*}_{q}&0&F_{-q}&Z_{\rm{B}}
\end{pmatrix}
\Psi_{q},
\end{align}
where $q$ is the crystal momentum along the chain direction, $\Psi^{\dagger}_{q}=(a^{\dagger}_{q},b^{\dagger}_{q},a_{-q},b_{-q})$ ($a_{q}$ and $b_q$ are the Fourier transforms of $a_i$ and $b_i$), and $F_q$, $G_{q}$, and $Z_{\eta}$ are given by
\begin{align}
F_q=&X_{q+}e^{iq}+Y_{q+}e^{-iq},\\
G_q=&X_{q-}e^{iq}+Y_{q-}e^{-iq},\\
Z_{\eta}=&-(1+\gamma)\left[J\cos(\theta_{\rm{A}}-\theta_{\rm{B}})+K\cos{\theta_{\rm{A}}}\cos{\theta_{\rm{B}}}\right]\nonumber\\
&+H\cos(\theta_{H}-\theta_{\rm{\eta}}),
\end{align}
with
\begin{align}
X_{q\pm}=&\frac{J}{2}\left\{\cos(\theta_{\rm{A}}-\theta_{\rm{B}})\pm 1\right\}+\frac{K}{2}\sin{\theta_{\rm{A}}}\sin{\theta_{\rm{B}}}\nonumber\\ 
&\mp \frac{i D}{2}\left(\cos\theta_{\rm{A}} \pm \cos\theta_{\rm{B}}\right),\\
Y_{q\pm}=&\gamma\left[\frac{J}{2}\left\{\cos(\theta_{\rm{A}}-\theta_{\rm{B}})\pm 1\right\}+\frac{K}{2}\sin{\theta_{\rm{A}}}\sin{\theta_{\rm{B}}} \right]\nonumber\\ 
&\pm \gamma' \frac{i D}{2}\left(\cos\theta_{\rm{A}} \pm \cos\theta_{\rm{B}}\right), 
\end{align}
where the subscript $\pm$ in $X_{q\pm}$ ($Y_{q\pm}$) represents the upper (lower) sign on the right-hand side.

\subsection{Result}
\label{result1}

Before presenting the magnon excitations, we briefly describe the two-sublattice ordering stabilized in the Hamiltonian in Eq.~\eqref{eq1}.
We consider two collinear magnetic states. 
One is the ferromagnetic state for nonzero $J<0$ and $H$ but $K=0$; the ferromagnetic moments are aligned in the magnetic field direction, i.e., $\theta_{\rm{A}}=\theta_{\rm{B}}=\theta_{H}$. 
We ignore the canting of magnetic moments in the presence of $D$ by supposing $D\ll |J| <H$, as discussed in Appendix~\ref{1DGS}. 
The other is that the collinear antiferromagnetic state with the staggered moments along the $z$ direction ($\theta_{\rm{A}}=0$ and $\theta_{\rm{B}}=\pi$), which is stabilized by $J>0$ and $K$ but $H=0$~\cite{PhysRevB.99.094401}. 
We take the parameter to satisfy $K\gtrsim (1-\gamma')^2D^2/[2(1+\gamma)^{2}J]$, which avoids the canting of the moments by $D$.
See also Appendix~\ref{1DGS} in details.

In the following, we show the nonreciprocal magnon excitations in Eq.~(\ref{eq3}) in the cases of the ferromagnetic state in Sec.~\ref{ferro1} and the antiferromagnetic state in Sec.~\ref{antiferro1}. 
We introduce the BMTD as a microscopic indicator for the nonreciprocal magnons in Sec.~\ref{toro}.

\subsubsection{Ferromagnetic state}
\label{ferro1}
In the ferromagnetic state ($G_{q}=0$ and $Z_{\rm{A}}=Z_{\rm{B}}\equiv Z$), the Bogoliubov Hamiltonian in Eq.~(\ref{eq3}) reduces to the $2\times 2$ matrix, whose eigenvalues are given by
 \begin{align} 
 \label{eq4}
 E^{\pm}_q=S[Z\pm |F_q|],
 \end{align}
where $E^{+}_q$ and $E^{-}_q$ represent the magnon bands for upper and lower branches, respectively.
We show the magnon dispersions at $J=-1$, $H=4$, $\theta_{H}=0$ and $\gamma=\gamma'=0.5$ for $D=0$ and $0.2$ in Fig.~\ref{fig1}(c).
The result clearly exhibits the asymmetric magnon dispersions in the presence of $D$. 
The degree of the asymmetry depends on $\theta_{H}$; a smaller $\theta_{H}$ leads to more asymmetric magnon dispersions and the magnon dispersions become symmetric for $\theta_{H}=\pi/2$, as shown in Fig.~\ref{fig1}(d).

The asymmetric band deformation indicates the appearance of the odd function of $q$ in the dispersions.
To investigate the behavior of the asymmetric band deformation, we perform the expansion in the limit of $q=0$, which is given by 
\begin{align}
 \label{eq51}
 E^{\pm}_q\sim &-S[J(1+\gamma)-H]\pm SJ(1+\gamma) \nonumber\\
 &\pm \frac{2 SD \cos{\theta_{H}}(\gamma+\gamma')}{1+\gamma}  q
 \mp \frac{2SJ\gamma}{1+\gamma}q^2
, \end{align}
where we consider the contributions up to $D/J$ and $q^2$.
The expression in Eq.~(\ref{eq51}) shows that the nonzero third term is a source of the asymmetric bottom shift.

There are three ingredients to determine the nature of the asymmetric band deformation in the ferromagnetic ordering. 
The first one is the DM interaction $D$, where large $D$ leads to larger nonreciprocity [Fig.~\ref{fig1}(c)].
The second one is the angle between the DM vector and the magnetic field $\theta_{H}$, which reaches at maxima for $\theta_{H}=0$ so that the ferromagnetic moment is parallel to the DM vector [Fig.~\ref{fig1}(d)]. 
The third one is the breathing parameter, $\Gamma \equiv (\gamma+\gamma')/(1+\gamma)$, whose behavior is plotted in Fig.~\ref{fig1}(e); $|\Gamma|$ becomes the largest at $\gamma'=1$ for arbitrary $\gamma$ or $\gamma=0$ and $\gamma'=-1$, and vanishes for $\gamma'=-\gamma$.

The factor $\Gamma$ in Eq.~(\ref{eq51}) appears owing to an interference between different magnon hopping processes, as shown in Fig~\ref{fig1}(f). 
For the magnon hopping process on the A-B-A bonds in Fig.~\ref{fig1}(f), the process consisting of the magnon hopping by the DM interaction on the A-B (B-A) bond and by the isotropic exchange interaction on the B-A (A-B) bond gives the contribution to the asymmetric band deformation. 
The total processes give the contribution as $(\gamma+\gamma') DJ$ (Note that $J$ is canceled out with the factor in the denominator), which explains the qualitative behavior of the asymmetric band deformations.

\subsubsection{Antiferromagnetic state}
\label{antiferro1}

In the collinear antiferromagnetic state ($F_{q}=0$ and $Z_{\rm{A}}=Z_{\rm{B}}\equiv Z$), the eigenvalues of the Bogoliubov Hamiltonian in Eq.~(\ref{eq3}) are given by
 \begin{align}
 \label{eq6}
 E^{\pm}_q=&S\sqrt{Z^2-|G_{\pm q}|^2}.
 \end{align}
The magnon dispersions at $J=1$, $K=0.1$, and $\gamma=\gamma'=0.5$ for $D=0$ and $D=0.2$ are shown in Fig.~\ref{fig1}(g). 
The magnon bands are shifted from the band bottom by introducing $D$, although the behavior is different from that in the ferromagnetic case in Fig.~\ref{fig1}(c); the former exhibits the symmetric band deformation with respect to $q$~\cite{PhysRevB.99.094401}, while the latter exhibits the antisymmetric
 one.
This is confirmed in the expansion with respect to $q$ in Eq.~\eqref{eq6}, which is expressed as
 \begin{align}
 \label{eq61}
 E^{\pm}_q\sim &S(1+\gamma)\sqrt{K(2J+K)}
 \mp  \frac{2SJD(\gamma+\gamma')
 }{(1+\gamma)\sqrt{K(2J+K)}}
 q\nonumber\\
 &+\frac{4S\gamma J^2}{(1+\gamma)\sqrt{K(2J+K)}}q^2, 
  \end{align}
up to $D/J$ and $q^2$. 
The linear $q$ term appears in the second term in Eq.~(\ref{eq61}), which is also accounted for by an interference among magnon hopping processes.
The signs of the coefficients of $q^2$, however, are different compared to the expression in Eq.~(\ref{eq51}), which results in a different way of the band deformation in the ferromagnetic and antiferromagnetic states.
Similar to the ferromagnetic case, the bottom shift owing to the linear $q$ term becomes the largest when the antiferromagnetic moment is parallel to the DM vector and vanishes when they are perpendicular to each other (not shown).

\subsubsection{Bond magnetic toroidal dipole}
\label{toro}

The results in Secs.~\ref{ferro1} and \ref{antiferro1} indicate that the magnon band bottoms are shifted in the presence of the DM interaction: The antisymmetric bottom shift appears in the ferromagnetic state and the symmetric bottom shift appears in the antiferromagnetic state. 
In addition, the relative angle between the magnetic moments and the DM vector is also an important factor, as seen in Eq.~(\ref{eq51}). 
From these observations, we introduce the following microscopic quantity as an indicator for the antisymmetric magnon shifted structure: 
\begin{align}
\label{MTD}
\mathbf{T}^{(ij)} \cdot \mathbf{\hat{r}}^{(ij)}= \mathbf{D}^{(ij)}\cdot \mathbf{M}^{(ij)}, 
\end{align}
where $\mathbf{M}^{(ij)}$ represents the averaged magnetic moments for $i$th and $j$th spins, $(\mathbf{S}_{i}+\mathbf{S}_{j})/2$ and $\mathbf{\hat{r}}^{(ij)}$ is the unit vector connecting $i$th and $j$th spins. 
$\mathbf{T}^{(ij)}$ represents the magnetic toroidal dipole on the bond $(ij)$ denoted as BMTD~\cite{Hayami_PhysRevB.101.220403,Hayami_PhysRevB.102.144441}.

The expression in Eq.~(\ref{MTD}) means that the nonreciprocal magnon bottom shift from $q=0$ occurs when the BMTD along the bond direction is activated, i.e., $\mathbf{D}^{(ij)}\cdot \mathbf{M}^{(ij)} \neq 0$.
For example, in the ferromagnetic state, the nonreciprocal magnon bottom shift occurs owing to $
\mathbf{T}^{(\rm AB)} \neq 0$ and $\mathbf{T}^{(\rm BA)}\neq 0$, while it does not appear in the antiferromagnetic state owing to $ \mathbf{T}^{(\rm AB)}=\mathbf{T}^{(\rm BA)}=0$. 
The emergence of the nonreciprocal bands by the BMTD is reasonable from a symmetry viewpoint, since the asymmetric magnon band structure is compatible with the presence of the time-reversal-odd polar vector, which corresponds to the magnetic toroidal dipole~\cite{Hayami_PhysRevB.102.144441}. 
The asymmetric band deformations owing to the magnetic toroidal dipole have also been discussed in the electronic systems~\cite{dubovik1990toroid,kopaev2009toroidal,Yanase_JPSJ.83.014703,Hayami_PhysRevB.90.024432}.

The concept of the BMTD in Eq.~(\ref{MTD}) can be applied to both polar and chiral magnetic systems without inversion symmetry. 
In polar magnets, the polar DM vector $\mathbf{D}^{{\rm polar} (ij)}$ on the bond $(ij)$ is related to the electric field $\mathbf{E}$ as $\mathbf{D}^{{\rm polar} (ij)} \propto \mathbf{\hat{r}}^{(ij)} \times\mathbf{E}$ [see also Fig.~\ref{fig1}(a)]. 
Then, one can find that the BMTD along the bond direction is related to $\mathbf{D}^{{\rm polar} (ij)} \cdot \mathbf{M}^{(ij)}$ as
\begin{align}
\label{eq7}
\mathbf{D}^{{\rm polar} (ij)}\cdot \mathbf{M}^{(ij)}&\propto ( \mathbf{\hat{r}}^{(ij)} \times\mathbf{E})\cdot \mathbf{M}^{(ij)} \sim \mathbf{T}^{(ij)} \cdot \mathbf{\hat{r}}^{(ij)},
\end{align}
where we use the relation $ \mathbf{E}\times \mathbf{M}^{(ij)} \sim \mathbf{T}^{(ij)}$~\cite{Spaldin_0953-8984-20-43-434203}.
Thus, $\mathbf{D}^{{\rm polar} (ij)} \cdot \mathbf{M}^{(ij)} \neq 0$ activates the BMTD, which describes the asymmetric magnon excitations, as discussed in Secs.~\ref{ferro1} and \ref{antiferro1}.

Meanwhile, in chiral magnets, the DM vector is parallel to the bond direction as $\mathbf{D}^{{\rm chiral} (ij)}\propto G_u \mathbf{\hat{r}}^{(ij)}$, where $G_u$ is the axial tensor with time-reversal even, which corresponds to the $3z^2-r^2$-type electric toroidal quadrupole where $\mathbf{\hat{z}}//\mathbf{\hat{r}}^{(ij)}$~\cite{hayami2018microscopic}.
Then, the BMTD along the bond direction is represented by $\mathbf{D}^{{\rm chiral} (ij)} \cdot \mathbf{M}^{(ij)}$ as
\begin{align}
\label{eq71}
\mathbf{D}^{{\rm chiral} (ij)}\cdot \mathbf{M}^{(ij)}&\propto G_u \mathbf{\hat{r}}^{(ij)} \cdot \mathbf{M}^{(ij)}\sim\mathbf{T}^{(ij)} \cdot \mathbf{\hat{r}}^{(ij)},
\end{align}
where we use the relation $G_u \mathbf{M}^{(ij)} \sim \mathbf{T}^{(ij)}$~\cite{hayami2018microscopic,PhysRevB.98.165110}.
Thus, the asymmetric magnon deformations can also be expected in chiral magnets once the BMTD is activated.

It is noted that the antiferromagnetic state does not exhibit the nonreciprocal magnon excitations because of $\mathbf{T}^{(\rm AB)}=\mathbf{T}^{(\rm BA)}=0$. 
Nevertheless, the symmetric band bottom shift is realized, as shown in Fig.~\ref{fig1}(g). 
This is presumably owing to nonzero $\mathbf{D}\cdot \mathbf{S}_{\rm A}$ and $\mathbf{D}\cdot \mathbf{S}_{\rm B}$. 
In other words, $\mathbf{D}^{(ij)}\cdot \mathbf{S}_i =-\mathbf{D}^{(ij)}\cdot \mathbf{S}_j$ leads to the antisymmetric
 band deformation depending on the bands but the symmetric band deformation in the whole bands. 
A similar situation have also been found in itinerant electron systems where the spin polarization depends on the sublattices but no spin polarizations in the whole systems, which is called the hidden spin polarization~\cite{zhang2014hidden,hayami2016emergent,PhysRevLett.118.086402,PhysRevB.102.085205,hayami2021spin}.

\section{Breathing kagome structure}
\label{kagome}
The concept of the BMTD in Sec.~\ref{toro} is generalized for a magnetic cluster in two- and three-dimensional systems. 
For any magnetic orderings, the BMTD can be evaluated at each bond, and as a result, the spatial distribution of the BMTD in a magnetic cluster is obtained, where we call it the CMT multipoles. 
The types of CMT multipole components depend on the magnetic and lattice structures. 
In this section, we show the application of the BMTD by examining the magnon excitations in the layered breathing kagome magnet. 
In Sec.~\ref{model2}, we show a spin model in a three-dimensional breathing kagome structure. 
We discuss when and how the nonreciprocal magnon excitations appear and their model parameter dependences in Sec.~\ref{result2}.
The Bogoliubov Hamiltonian under three-sublattice orderings obtained by the Holstein-Primakov transformation is shown in Appendix~\ref{SW2}.

\subsection{Spin model}
\label{model2}

\begin{figure}[h]
\begin{center}
\includegraphics[width=1.00\linewidth]{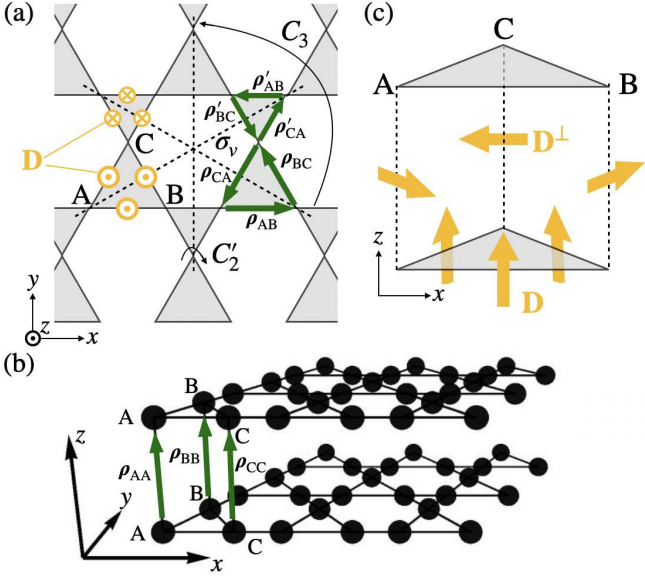}
\end{center}
\caption{(a) The two-dimensional plane in the layered breathing kagome structure. 
A part of point group symmetries ($C_3$, $C'_2$, and $\sigma_{\rm v}$) are also shown.
(b) The stacking of the breathing kagome structure along the $z$ direction.
(c) The nonzero DM vectors denoted as $D$ and $D^{\perp}$. 
}
\label{fig2}
\end{figure}

We consider a layered breathing kagome structure without inversion symmetry.  
The two-dimensional plane consists of upward and downward triangles with different sizes, as shown in Fig.~\ref{fig2}(a).
The two-dimensional planes are stacked along the $z$ direction, as shown in Fig.~\ref{fig2}(b). 
The present system includes three sublattices in the unit cell denoted as A, B, and C, whose symmetry is characterized by the point group $\rm{D_{3h}}$ consisting of the symmetry operations with respect to the threefold rotation around the $z$ axis $C_3$, twofold rotation around the $y$ axis $C'_2$, the mirror in the $xy$ and $xz$ planes $\sigma_{h}$ and $\sigma_{v}$, and threefold rotoreflection around the $z$ axis, as shown in Fig.~\ref{fig2}(a).
We define the bond vectors shown in Figs.~\ref{fig2}(a) and \ref{fig2}(b) as
\begin{align}
\label{bond1}
&\boldsymbol{\rho}_{\rm{AB}}=-\boldsymbol{\rho}'_{\rm{AB}}=(1,0,0),\\
&\boldsymbol{\rho}_{\rm{BC}}=-\boldsymbol{\rho}'_{\rm{BC}}=\left(-\frac{1}{2},\frac{\sqrt{3}}{2},0\right),\\
&\boldsymbol{\rho}_{\rm{CA}}=-\boldsymbol{\rho}'_{\rm{CA}}=\left(-\frac{1}{2},-
\frac{\sqrt{3}}{2},0\right),\\
\label{bond2}
&\boldsymbol{\rho}_{\rm{AA}}=\boldsymbol{\rho}_{\rm{BB}}=\boldsymbol{\rho}_{\rm{CC}}=(0,0,1),
\end{align}
where we take the norm of $\boldsymbol{\rho}'_{\eta \eta'}$ and $\boldsymbol{\rho}_{\eta \eta'}$ unity, and their difference is expressed as the different magnitude of the interactions.

The spin model is constructed to satisfy above point group symmetries.
The Hamiltonian on the two-dimensional plane, $H^{\rm intra-plane}$, is given by
\begin{align}
\label{eq9}
&H^{\rm intra-plane}=
\sum_{\langle i\eta, j \eta' \rangle}
\sum_{\alpha, \beta } 
\sum_{\rm{P}}
S^{\alpha}_{i\eta} J^{{\rm{P}}\alpha \beta}_{\eta \eta'} S^{\beta}_{j\eta'},
\end{align}
where $S^{\alpha}_{i\eta}$ is the $\alpha$ component of the classical spin at site $i$ for sublattice $\eta=$ A, B, and C.
The sum of ${\langle i\eta, j \eta' \rangle}$ runs all bonds including upward and downward triangles
The interactions in the upward triangle $J^{\triangle}_{\eta \eta'}$ and the downward triangle $J^{\bigtriangledown}_{\eta \eta'}$
 are distinguished by the symbol $\rm{P}=\bigtriangledown, \triangle$.
The components of the interaction tensors are given by
\begin{align}
\label{eq10}
&J^{{\rm{P}}}_{\rm{AB}}=
\begin{pmatrix}
\xi({\rm{P}})(J+J^a)&\xi'({\rm{P}})D&0\\
-\xi'({\rm{P}})D&\xi({\rm{P}})(J-J^a)&0\\
0&0&\xi({\rm{P}})J^z
\end{pmatrix},\\
\label{eq11}
&\left(J^{\rm{P}}_{\rm{BC}},J^{\rm{P}}_{\rm{CA}}\right)=
\big(R(C_3) J^{\rm{P}}_{\rm{AB}}R^{-1}(C_3), R(C_3^2)J^{\rm{P}}_{\rm{AB}}R^{-1}(C_3^2)\big),
\end{align}
where $\xi(\triangle)=\xi'(\triangle)=1$, $\xi(\bigtriangledown)=\gamma$ and $\xi'(\bigtriangledown)=\gamma'$.
The interactions of the AB bond given in Eq.~\eqref{eq10} include the $xy$- and $z$-spin exchange interactions $J$ and $J^z$, the bond-dependent anisotropic exchange interaction $J^a$, and the DM interaction $D$.
The interactions on the BC and CA bonds in Eq.~\eqref{eq11} are connected with those on the AB bond by the threefold rotational matrices $R(C_3)$ and $R(C_3^2)$. 
$\gamma$ and $\gamma'$ ($0<\gamma, \gamma' \leq 1$) are the breathing parameters between upward and downward triangles for $(J, J^a, J^z)$ and $D$, respectively.

The inter-layer Hamiltonian connecting the two-dimensional planes, $H^{\rm inter-plane}$,
is given by
\begin{align}
\label{eq9-2}
&H^{\rm inter-plane}=
\sum_{\langle i\eta, j \eta \rangle}
\sum_{\alpha, \beta } 
S^{\alpha}_{i\eta} J^{\perp \alpha \beta}_{\eta \eta} S^{\beta}_{j \eta}.
\end{align}
The sum of ${\langle i\eta, j \eta \rangle}$ runs all the nearest-neighbor bonds along the $z$ direction.
The components of the interaction tensors $J^{\perp}_{\eta \eta}$ are given by
\begin{align}
\label{eq12}
&J^{\perp}_{\rm{CC}}=
\begin{pmatrix}
J^{\perp }+J^{\perp a}&0&0\\
0&J^{\perp }-J^{\perp a}&D^{\perp}\\
0&-D^{\perp}&J^{\perp z}
\end{pmatrix},  \\
\label{eq13}
&\left(J^{\perp}_{\rm{AA}}, J^{\perp}_{\rm{BB}}\right)=\left(R(C_3)J^{\perp}_{\rm{CC}}R^{-1}(C_3), R(C_3^2)J^{\perp}_{\rm{CC}}R^{-1}(C_3^2)\right).
\end{align}
The interactions for A-A, B-B, and C-C bonds in Eq.~\eqref{eq12} include the $xy$- and $z$-spin exchange interactions $J^{\perp}$ and $J^{\perp z}$, the bond-dependent symmetric anisotropic exchange interaction $J^{\perp a}$,  and the bond-dependent DM interaction $D^{\perp}$.
The direction of the DM vector is different from that in Eq.~(\ref{eq10}): The former is lied on the $xy$ plane and the latter is along the $z$ direction, as schematically shown in Figs.~\ref{fig2}(a) and \ref{fig2}(c).
Moreover, it is noted that the DM interaction $D^{\perp}$ is present only for the breathing structure, i.e. $D^{\perp}=0$ for $\gamma=\gamma'=1$.

\subsection{Result}
\label{result2}

We first discuss the classical ground state of the model $H=H^{\rm{inter-plane}}+H^{\rm{intra-plane}}$ within the three-sublattice orderings.
There are nine independent magnetic structures, which are classified by the irreducible representation  $\Gamma^{\rm{irrep}}$ of $\rm{D_{3h}}$.
For the bases $\mathbf{M}^{\Gamma^{\rm{irrep}}}=(\mathbf{S}^{\Gamma^{\rm{irrep}}}_{\rm{A}},\mathbf{S}^{\Gamma^{\rm{irrep}}}_{\rm{B}},\mathbf{S}^{\Gamma^{\rm{irrep}}}_{\rm{C}})$, the three-sublattice spin patterns are given by
\begin{subequations}
\label{eq15}
\begin{align}
\label{a''1}
&\mathbf{S}^{\rm{A''_1}}_{\eta}=S\mathbf{r}_{\eta},\\
\label{a'2}
&\mathbf{S}^{\rm{A'_2}}_{\eta}=S(0,0,1),\\
\label{a''2}
&\mathbf{S}^{\rm{A''_2}}_{\eta}=S(-y_{\eta},x_{\eta},0),\\
\label{e'}
&\begin{cases}
\mathbf{S}^{\rm{E'_1}}_{\eta}=\sqrt{2}S(0,0,-x_{\eta})\\
\mathbf{S}^{\rm{E'_2}}_{\eta}=\sqrt{2}S(0,0,-y_{\eta})
\end{cases},\\
\label{e''1}
&\begin{cases}
\mathbf{S}^{\rm{E''_{\rm{AFM}1}}}_{\eta}=S(x^2_{\eta}-y^2_{\eta}, 2x_{\eta}y_{\eta},0)\\
\mathbf{S}^{\rm{E''_{\rm{AFM}2}}}_{\eta}=S(-2x_{\eta}y_{\eta}, x^2_{\eta}-y^2_{\eta},0)
\end{cases},\\
\label{e''2}
&\begin{cases}
\mathbf{S}^{\rm{E''_{\rm{FM}1}}}_{\eta}=S(1,0,0)\\
\mathbf{S}^{\rm{E''_{\rm{FM}2}}}_{\eta}=S(0,1,0)
\end{cases}
,\end{align}
\end{subequations}
where $(x_{\rm{A}}, y_{\rm{A}})=(-\sqrt{3}/2,-1/2)$, $(x_{\rm{B}}, y_{\rm{B}})=(\sqrt{3}/2,-1/2)$, and $(x_{\rm{C}}, y_{\rm{C}})=(0,1)$.
We identify the classical ground state by minimizing the energy for the above bases, as detailed in Appendix~\ref{GS2}.
We show the Bogoliubov Hamiltonian under the three-sublattice orderings in Appendix~\ref{SW2}.

In the following, we discuss the magnon spectra in four magnetic ordered states in Sec.~\ref{result2} and two ones in Appendix~\ref{other}, which are stabilized as the classical ground state by appropriately choosing the model parameters.
We show various angle dependences of nonreciprocal magnons depending on the magnetic ordered states and the DM interactions $D$ and $D^{\perp}$. 
To identify the angle dependences of asymmetric magnon dispersions, we perform the series expansion in terms of $\mathbf{q}=(q_x, q_y, q_z)=q(\sin{\theta}\cos{\phi}, \sin{\theta}\sin{\phi}, \cos{\theta})$ for $q \to 0$, which is represented by  
\begin{align}
\label{28}
E_{\mathbf{q}}=&a_0+q\big[\sin{\theta}(a_{x}\cos{\phi} +a_{y}\sin{\phi}) +a_{z}  \cos{\theta}\big]\nonumber\\ 
&+q^2\big[\sin^2{\theta} (a_{xy}\sin 2\phi +a_v \cos2\phi) \nonumber\\ 
&+\sin{\theta}\cos{\theta}(a_{zx}\cos{\phi}+a_{yz} \sin{\phi})\nonumber\\
&+a_{u}(3\cos^2{\theta}-1)\big]\nonumber\\
&+q^3\big[\sin^3\theta ( a_{3a} \cos3\phi+a_{3b}\sin 3\phi)\nonumber\\
&+\cos{\theta}\sin^2{\theta} (a_{z}^{\beta} \cos 2\phi+a_{xyz} \sin 2\phi)\nonumber\\
&+\sin{\theta}(5\cos^2{\theta}-1)(a_{3u}\cos{\phi}+a_{3v}\sin{\phi})\nonumber\\
&+a_{z}^{\alpha}\cos{\theta}(5\cos^2{\theta}-3)\big]+O(q^4)
,\end{align}
where $a_0$, $(a_x, a_y, a_z)$, $(a_u, a_{v}, a_{yz}, a_{zx}, a_{xy})$, and $(a_{3a}, a_{3b}, a_{z}^{\beta}, a_{xyz}, a_{3u},  a_{3v}, a_{z}^{\alpha})$ represent the expansion coefficients in the order of 0, 1, 2, and 3 in terms of $q$, respectively.
As the odd order of $q$ corresponds to the active magnetic toroidal multipoles~\cite{PhysRevB.98.165110},
one can expect what types of magnetic toroidal multipoles become active by calculating the coefficients in Eq.~(\ref{28}).

\begin{figure}[h]
\begin{center}
\includegraphics[width=0.92\linewidth]{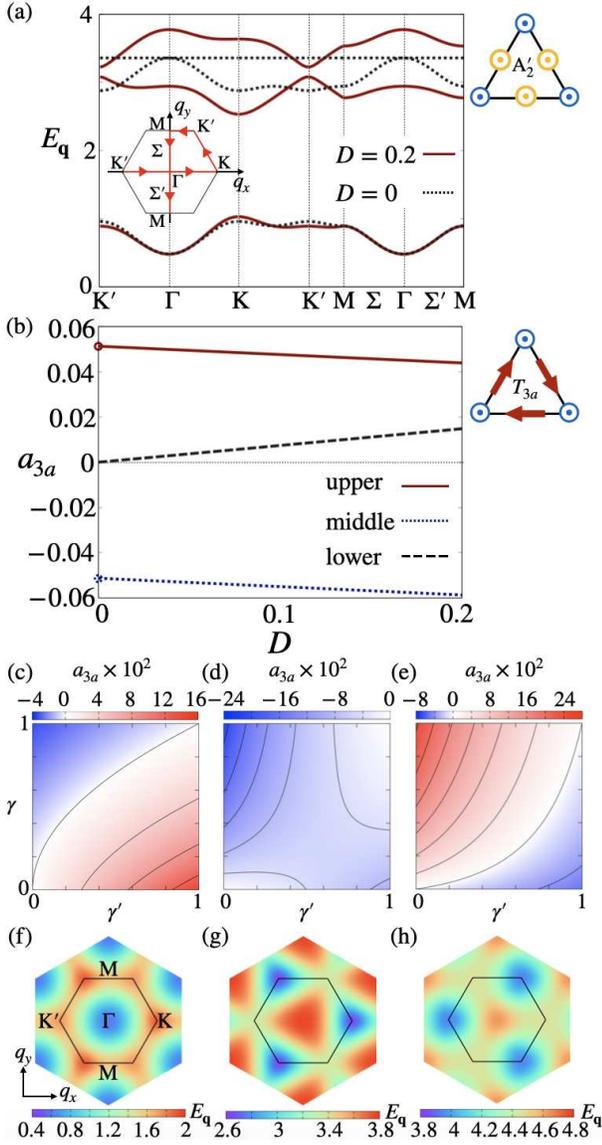}
\end{center}
\caption{
(a) Magnon dispersions in the $\mathbf{M}^{\rm{A'_2}}$ state along the high symmetrical lines (see the inset).  
The spin configuration is shown in the right panel.
The results for $D=0$ and $D=0.2$ are shown at $(J, J^z \gamma, \gamma' )=(-0.8, -1, 0.2, 0.2)$ (other parameters are zero).
(b) $D$ dependence of $a_{3a}$ for three branches in (a). 
The schematic picture of the CMT octupole $T_{3a}$ corresponding to nonzero $a_{3a}$ is shown in the right panel.
(c)-(e) Intensity plot of $a_{3a}$ while varying $\gamma$ and $\gamma'$ for (c) lower, (d) middle, and (e) upper branches. 
Contour lines are drawn every $0.04$.
(f)-(h) The color plot of magnon dispersions for (f) lower, (g) middle, and (h) upper  branches calculated at $\gamma=0.5$ and $\gamma'=0.25$.
}
\label{fig3}
\end{figure}

\subsubsection{Ferromagnetic state with the moments along the $z$ axis }
\label{a2}
We consider the ferromagnetic state with the moments along the $z$ direction ($\mathbf{M}^{\rm{A'_2}}$), which belongs to the $\rm{A'_2}$ irreducible representation.
This state is stabilized when the intra-plane interactions are given by $0<D\ll-J<-J^z$, as shown in Appendix~\ref{GS2}. 
We ignore the irrelevant parameters for the nonreciprocal magnons, i.e., $J^a=J^{\perp}=J^{\perp a}=J^{\perp z}=D^{\perp}=0$, and then it is enough to consider the two-dimensional lattice structure with $\theta=\pi/2$ in Eq.~(\ref{28}).

Figure~\ref{fig3}(a) shows the magnon dispersions at $(J, D, J^z, \gamma, \gamma' )=(-0.8, 0.2, -1, 0.2, 0.2)$ (red solid 
curves).
For comparison, we also show the magnon dispersions at $D=0$ by the black dotted curves.
Compared to the result at $D=0$, the magnon dispersions at $D=0.2$ are asymmetrically inclined on the $\rm{K'}$-$\rm{\Gamma}$-$\rm{K}$ line not on the M-$\Gamma$ line.
This implies that the angle dependence of the nonreciprocal magnons is represented by $\cos 3\phi$. 
Indeed, by calculating the coefficients in Eq.~(\ref{28}), we obtain a nonzero $a_{3a}$ for the functional form of $q_x(q_x^2-3q_y^2)$.
As shown in Fig.~\ref{fig3}(b), $a_{3a}$ behaves linearly while increasing $D$.
Note that a finite jump of the upper and middle branches for an infinitesimally small $D$ is owing to the gap opening at the $\Gamma$ point.

The emergence of the nonreciprocal magnons is interpreted as the BMTD introduced in Sec.~\ref{toro}. 
By evaluating the BMTD in Eq.~\eqref{MTD} for each bond in the breathing kagome structure, we obtain a swirling structure of the BMTD, as shown in the right panel of Fig.~\ref{fig3}(b). 
This spatial distribution of the BMTDs in the unit cell corresponds to the CMT octupole named $T_{3a}$, which is compatible with nonzero $a_{3a}$~\cite{PhysRevB.99.174407}.

The asymmetric coefficient $a_{3a}$ also depends on the breathing parameter similar to the one-dimensional case in Sec.~\ref{ferro1}, as shown in Figs.~\ref{fig3}(c)-\ref{fig3}(e), where the results for the lower, middle, and upper branches are shown in Figs.~\ref{fig3}(c), \ref{fig3}(d), and \ref{fig3}(e), respectively. 
In all the branches in Figs.~\ref{fig3}(c)-\ref{fig3}(e), $|a_{3a}|$ tends to be larger for $\gamma=0$ and $\gamma'=1$ or $\gamma=1$ and $\gamma'=0$, which indicates that the inequivalence of the breathing parameters for the symmetric and antisymmetric exchange interactions, $\gamma$ and $\gamma'$, is important in realizing the large nonreciprocal magnon dispersions. 
Meanwhile, $a_{3a}$ in the lower branch in Fig.~\ref{fig3}(c) vanishes when $\gamma'=\gamma^2$, which implies that there are no asymmetric modulations in the limit of $q \to 0$. 
Nevertheless, it is noted that asymmetric magnon dispersions are found far from $\mathbf{q}=\mathbf{0}$, 
as shown in Fig.~\ref{fig3}(f), where the color plot represents the magnon dispersion for the lower branch calculated at $(J, D, J^z, \gamma, \gamma' )=(-0.8, 0.2, -1, 0.5, 0.25)$ satisfying $\gamma'=\gamma^2$. 
We also plot the magnon dispersions for the middle and higher branches for a reference in Figs.~\ref{fig3}(g) and \ref{fig3}(h), which also show nonzero $a_{3a}$. 
The results mean that the perturbative expansion in terms of $q$ sometimes fails to describe the nonreciprocal magnons apart from $\mathbf{q}=\mathbf{0}$.
Meanwhile, the concept of the CMT multipoles can be applied irrespective of the model parameters and it is easy to predict the angle dependence of the nonreciprocal magnons based on the bond degree of freedom.

\begin{figure}[h]
\begin{center}
\includegraphics[width=0.92\linewidth]{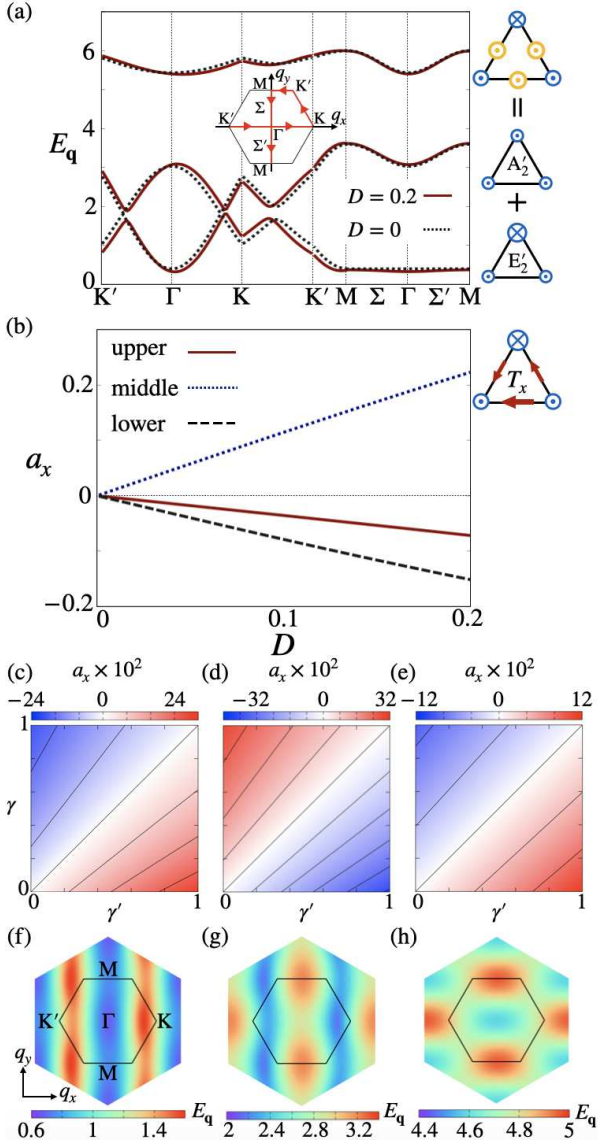}
\end{center}
\caption{
(a) Magnon dispersions in the $\mathbf{M}^{++-}$ state along the high symmetrical line (see the inset).
The spin configuration is shown in the right panel.
The results are calculated at $(J, J^z, J^{\perp z}, \gamma, \gamma' )=(0.8, 1, -1, 1, 0.2 )$ for $D=0$ and $D=0.2$ (other parameters are zero).
(b) $D$ dependence of $a_{x}$ for the branches in (a).
The schematic picture of the CMT dipole $T_{x}$ corresponding to nonzero $a_{x}$ is shown in the right panel.
(c)-(e) Intensity plot of $a_{x}$ while varying $\gamma$ and $\gamma'$ for (c) lower, (d) middle, and (e) upper branches. 
Contour lines are drawn every $0.08$ for (c) and (d), and $0.04$ for (e).
(f)-(h) The color plot of magnon dispersions for (f) lower, (g) middle, and (h) upper  branches calculated at $\gamma=\gamma'=0.5$.
}
\label{fig4-1}
\end{figure}

\subsubsection{Up-up-down state }
\label{e}
We consider the up-up-down structure $\mathbf{M}^{++-}$ given in Eq.~\eqref{eq221} in Appendix~\ref{GS2}, which is constructed by the superpositions of the $\rm{A'_{2}}$ and $\rm{E'}_2$ states, as shown in the right panel in Fig.~\ref{fig4-1}(a).
In order to stabilize the $\mathbf{M}^{++-}$ state as a meta-stable state, we open the magnon band gap by introducing $J^{\perp z}$ besides the condition of $0<D\ll J<J^z$.
The other parameters $J^a$, $J^{\perp}$, $J^{\perp a}$, and $D^{\perp }$ are set to be zero.

Figure~\ref{fig4-1}(a) shows the magnon dispersions at $(J, J^z, D, J^{\perp z}, \gamma, \gamma' )=(0.8, 1, 0.2,  -1, 1, 0.2 )$ (red solid curves) in the $\mathbf{M}^{++-}$ state.
Compared to the black dotted curves at $D=0$, the bottom of magnon dispersions at $D=0.2$ shifts along the $\rm{K'}$-$\rm{\Gamma}$-$\rm{K}$ line. 
This implies that the angle dependences of the nonreciprocal magnons are represented by $\cos{\phi}$, which are confirmed by calculating $a_x$ in Eq.~(\ref{28}). 
As shown in Fig.~\ref{fig4-1}(b), $a_{x}$ behaves linearly while increasing $D$.

By evaluating the BMTD in Eq.~\eqref{MTD} similar to Sec.~\ref{a2}, we obtain the BMTD configuration to have the uniform component along the $x$ direction, as shown in the right panel of Fig.~\ref{fig4-1}(b). 
Thus, the $x$ component of the magnetic toroidal dipole is activated in the cluster~\cite{PhysRevB.99.174407}, which is consistent with the results in Fig.~\ref{fig4-1}(a).

The coefficient $a_{x}$ depends on the breathing parameter, as shown in Figs.~\ref{fig4-1}(c)-\ref{fig4-1}(e). 
In all the branches in Figs.~\ref{fig4-1}(c)-\ref{fig4-1}(e), $|a_{x}|$ tends to be larger for $\gamma=0$ and $\gamma'=1$ or $\gamma=1$ and $\gamma'=0$. 
The coefficient $a_{x}$ in all the branches vanishes when $\gamma'=\gamma$, although the asymmetric modulations are still found far from $\mathbf{q}=\mathbf{0}$, as shown in Figs.~\ref{fig4-1}(f)-\ref{fig4-1}(h), similar to the ferromagnetic case in Sec.~\ref{a2}.

\begin{figure}[h]
\begin{center}
\includegraphics[width=0.92\linewidth]{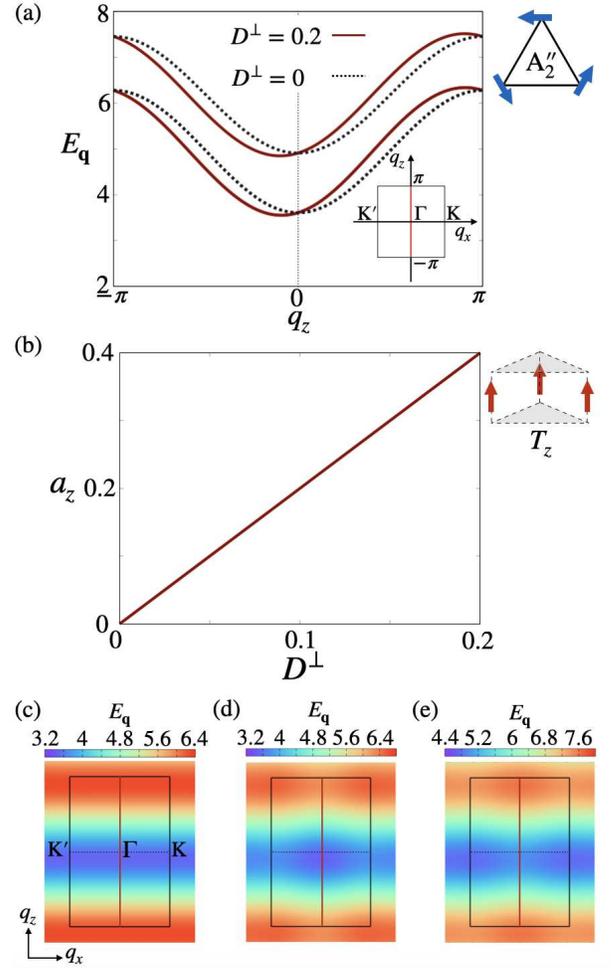}
\end{center}
\caption{
(a) Magnon dispersions in the $\mathbf{M}^{\rm{A''_2}}$ state along the $z$ axis (see the inset).  
The spin configuration is shown in the right panel.
The results for $D^{\perp}=0$ and $D^{\perp}=0.2$ are shown at  $(J, J^a, J^z, D, J^{\perp}, J^{\perp a}, J^{\perp z}, \gamma, \gamma')=(1, -0.5, 0.8, 0, -1, -0.5, -0.8,  0.2, 0.2)$.
It is noted that magnon dispersions for lower and middle branches are degenerate at $q_x=0$.
(b) $D$ dependence of $a_{z}$ for three branches in (a), where all the branches show the same behavior.
The schematic picture of the CMT dipole $T_{z}$ is shown in the right panel.
(c)-(e) The color plot of magnon dispersions for (f) lower, (g) middle, and (h) upper  branches.
}
\label{fig5}
\end{figure}

\subsubsection{Noncollinear antiferromagnetic state}
\label{a22}

We consider the antiferromagnetic ($\mathbf{M}^{\rm{A''_2}}$) state with the swirling spin configuration in the $xy$ plane.
This spin configuration belongs to the $\rm{A''_2}$ irreducible representation, which is stabilized for $0<J^z<J$ and $J^a<0$, as shown in Appendix~\ref{GS2}. 
For the inter-plane interactions, we take $0<D^{\perp}\ll -J^{\perp z}<-J^{\perp}$ and $J^{\perp a}<0$.
We here take $D=0$ for simplicity.

Figure~\ref{fig5}(a) shows the magnon dispersions at $(J, J^a, J^z, J^{\perp}, J^{\perp a}, J^{\perp z}, D^{\perp},  \gamma, \gamma')=(1, -0.5, 0.8, -1, -0.5, -0.8, 0.2,   0.2, 0.2)$ (red solid curves).
Compared to the result at $D^{\perp}=0$ (black dotted curves), the magnon dispersions at $D^{\perp}=0.2$ are asymmetrically shifted along the $z$ direction with the angle dependence of $\cos{\theta}$, indicating the nonzero $a_z$.
The coefficient $a_z$ is proportional to $D^{\perp}$, as shown in Fig.~\ref{fig5}(b).
The color plots of magnon dispersions are shown in Figs.~\ref{fig5}(c)-\ref{fig5}(e). 

By evaluating the BMTD in Eq.~\eqref{MTD} for each bond connecting to the breathing kagome planes, we obtain the ferroic alignment of the BMTD, as shown in the right panel of Fig.~\ref{fig5}(b). 
This corresponds to the emergence of the CMT dipoles along the $z$ direction $T_{z}$~\cite{PhysRevB.99.174407}.

\begin{figure}[t]
\begin{center}
\includegraphics[width=0.88\linewidth]{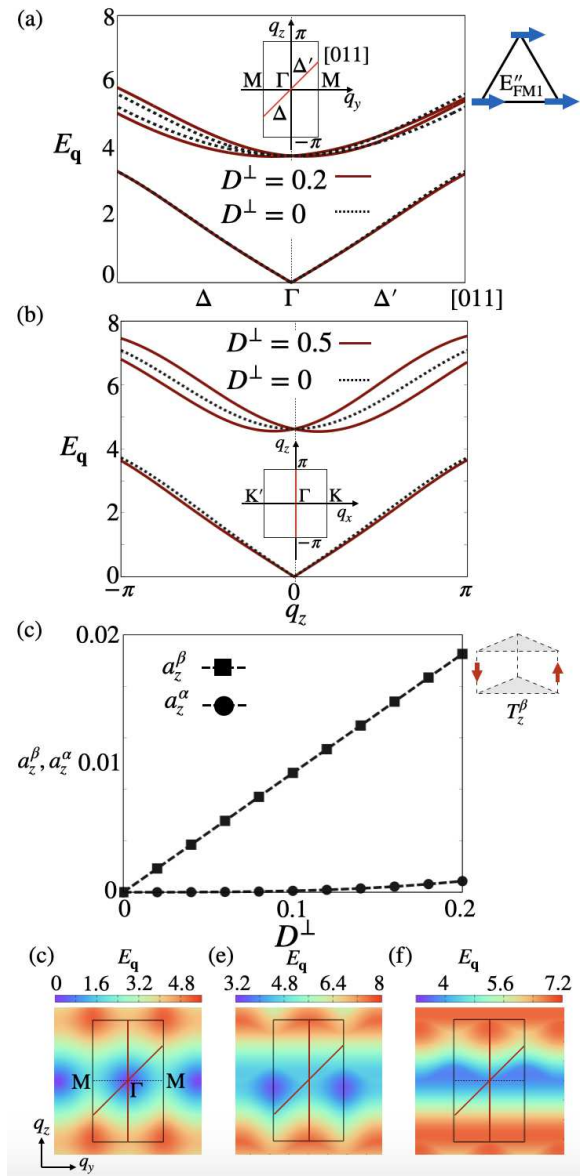}
\end{center}
\caption{
Magnon dispersions in the $\mathbf{M}^{\rm{E''_{FM 1}}}$
 state along (a) the [011] line and (b) the $z$ axis (see the inset).  
The spin configuration is shown in the right panel in (a).
The results for (a) $D^{\perp}=0$ and $D^{\perp}=0.2$ and (b) $D^{\perp}=0$ and $D^{\perp}=0.5$ are shown at $(J, J^z, J^{\perp}, J^{\perp z}, \gamma)=(-1, -0.5, -1, -0.5, 0.5)$ (other parameters are zero).
(c) $D^{\perp}$ dependence of $a^{\beta_{z}}$ (square symbol) and $a^{\alpha}_{z}$ (round symbol) for the lower branches in (a) and (b).
The schematic picture of the CMT dipole $T^{\beta}_{z}$ corresponding to nonzero $a^{\beta}_{z}$ is shown in the right panel.
(d)-(f) The color plot of magnon dispersions for (f) lower, (g) middle, and (h) upper  branches.
}
\label{fig6-2}
\end{figure}

\subsubsection{Ferromagnetic state with the moments along the $x$ axis }
\label{e2}

We consider the ferromagnetic state with the moments along the $x$ direction ($\mathbf{M}^{\rm{E''_{FM1}}}$
), which belongs to the $\rm{E''}$ irreducible representation.
This state is stabilized when the intra-plane interactions are given by $0<-J^z<-J$ and $J^a=D=0$, as shown in Appendix~\ref{GS2}. 
We also introduce the inter-plane interactions as $0<D^{\perp} \ll -J^{\perp z}<-J^{\perp }$ but $J^{\perp a}=0$.

Figures~\ref{fig6-2}(a) and \ref{fig6-2}(b) show the magnon dispersions at $D^{\perp}=0.2$ along the [011] line and at $D^{\perp}=0.5$ along the $z$ direction, respectively (red solid curves).
Other parameters are given by $(J, J^z, J^{\perp}, J^{\perp z}, \gamma)=(-1, -0.5, -1, -0.5, 0.5)$.
Compared to the result at $D^{\perp}=0$ (black dotted curves), the magnon dispersions are asymmetrically inclined in both the [011] and
$z$ directions.
The angle dependences of the nonreciprocal magnons are represented by $\cos{\theta}\sin^2{\theta}\cos 2\phi$ and $\cos^3{\theta}$, which correspond to nonzero $a^{\beta}_{z}$ and $a^{\alpha}_z$, respectively. 
The behaviors of $a^{\beta}_{z}$ and $a^{\alpha}_{z}$ are different as shown in Fig.~\ref{fig6-2}(c); 
$a^{\beta}_{z}$ ($a^{\alpha}_{z}$) linearly (third-order nonlinearly) increases against $D^{\perp}$. 

The emergence of the nonreciprocal magnons is also accounted for by the BMTD.
In the present case, the staggered alignment of the BMTD is obtained, as shown in the right panel of Fig.~\ref{fig6-2}(c), which corresponds to the appearance of the CMT octupole $T^{\beta}_{z}$~\cite{PhysRevB.99.174407}. 
The large coefficient $a^{\beta}_{z}$ is related to $T^{\beta}_{z}$, while the small one $a^{\alpha}_{z}$ is related to the other magnetic toroidal octupole $T^{\alpha}_{z}$, which is secondary activated in the $\mathbf{M}^{\rm{E''_{FM}2}}$ ordering, since $T^{\alpha}_{z}$ belongs to the same irreducible representation as $T^{\beta}_{z}$~\cite{yatsushiro2021multipole}.
Nonzero $T^{\alpha}_{z}$ is understood from the higher-order contribution of the BMTD, which will be reported elsewhere.

\begin{figure}[h]
\begin{center}
\includegraphics[width=1.00\linewidth]{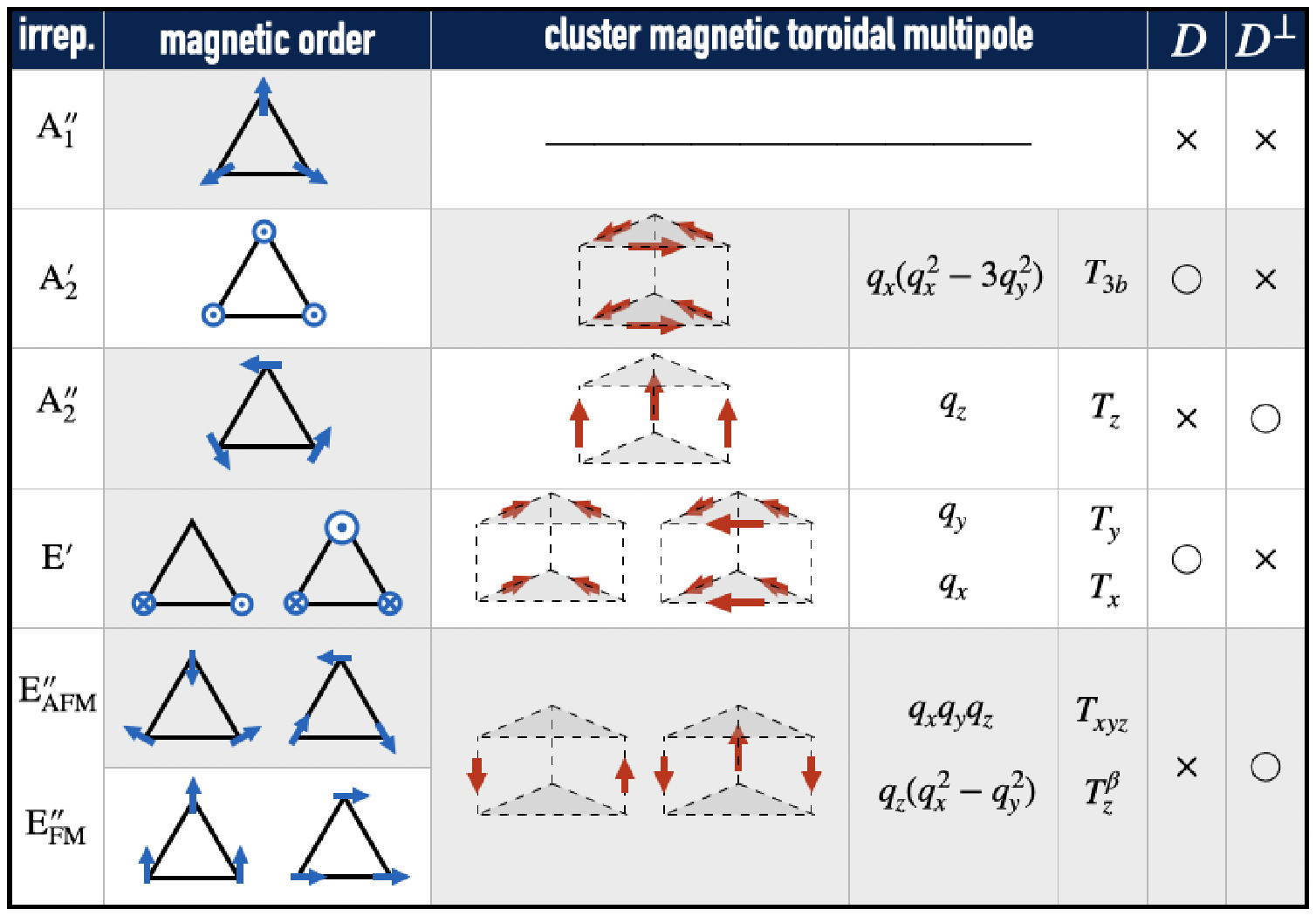}
\end{center}
\caption{
Classifications of nine three-sublattice spin configurations and six CMT multipoles with respect to the irreducible representation (irrep.) of the point group $\rm{D_{3h}}$. 
The corresponding asymmetric magnon dispersions are also shown.
The rightmost two columns represents the key DM interactions for the nonreciprocal magnons, where $\bigcirc$ means the necessary DM interaction. 
}
\label{fig7}
\end{figure}

\subsection{Summary of cluster magnetic toroidal multipoles in the breathing kagome system}
\label{dis}

We summarize the relation between the DM interactions, magnetic orderings, and CMT multipoles in each irreducible representation of the breathing kagome system in Fig~\ref{fig7}.
One can easily find when and how the nonreciprocal magnons appear by evaluating the BMTD at each bond in the magnetic cluster.
Once one of the CMT multipoles is activated, the magnon bands are deformed in an asymmetric way, whose angle dependences are characterized by the types of multipoles~\cite{PhysRevB.98.165110}. 
As the concept of the CMT multipoles consisting of the BMTD is generic, it can be applied to magnetic orderings under any lattice structures when a magnetic unit cell is the same as a lattice one. 

In this way, the classification of the nonreciprocal magnons by the CMT multipoles gives key model parameters, which are not explained by the magnetic point-group analysis. 
In other words, one can find which DM interactions in the system play an important role in realizing the large nonreciprocity by calculating the BMTD at each bond.  
Our results provide a guiding principle to obtain the large antisymmetric dispersions from the microscopic viewpoint.

\section{Summary}
\label{summary}

To summarize, we have investigated the conditions for the nonreciprocal magnon excitations in noncentrosymmetric magnets with the DM interaction.
By analyzing the one-dimensional spin model with the DM interaction, we obtained the concept of the BMTD in Eq.~\eqref {MTD}, which are defined by the inner product between the DM vector and the averaged spin moments at the ends of the bonds.
We applied the concept of the BMTD to the collinear and noncollinear spin orderings in the breathing kagome structure.
We have demonstrated that the nonreciprocal magnon excitations are well described by the CMT multipoles.
Our study will stimulate a further exploration of nonreciprocal magnon physics, which is expected to be realized in the materials with the breathing kagome structure, such as $\rm{(NH_4)_2[C_7H_{14}N][V_7O_6F_{18}]}$~\cite{PhysRevLett.118.237203} and $\rm{Li_2In_{1-x}Sc_xMo_3O_8}$~\cite{PhysRevLett.120.227201}. 
The other noncentrosymmetric magnets, such as Nd$_5$Si$_4$~\cite{cadogan2002magnetic}, Ho$_2$Ge$_2$O$_7$~\cite{morosan2008structure, playford2017situ, hallas2012statics}, YMnO$_3$~\cite{abrahams2001ferroelectricity, fujimura1996epitaxially}, and U$_3$As$_4$~\cite{burlet1981non}, are also candidate materials in the present scheme.
Furthermore, the concept of the BMTD can be applicable to the centrosymmetric lattice structures, where the inversion symmetry is preserved globally but broken locally at atomic sites, since there is a sublattice-dependent DM interaction in the system~\cite{doi:10.7566/JPSJ.85.053705}. 
In this situation, the antiferromagnetic ordering which breaks the spatial inversion symmetry can give rise to the nonreciprocal magnon.
The candidate materials belonging to this category, for example, are the honeycomb antiferromagnet $\rm{ErNi_3Ga_9}$~\cite{e1ddc90bb22e4448a040d6ce3fdca500} and $\rm{Co_4Nb_2O_9}$~\cite{PhysRevB.93.075117, PhysRevB.97.085154,Yanagi2017,Yanagi_PhysRevB.97.020404,matsumoto2019symmetry}.

\begin{acknowledgments}
This research was supported by JSPS KAKENHI Grants Numbers JP18K13488, JP19K03752, JP19H01834, JP21H01037, and by JST PRESTO (JPMJPR20L8). 
Parts of the numerical calculations were performed in the supercomputing systems in ISSP, the University of Tokyo.
\end{acknowledgments}

\appendix

\section{Two-sublattice orderings in the one-dimensional spin chain}
\label{1DGS}

In this appendix, we investigate a stable spin configuration of the model in Eq.~\eqref{eq1} within the two-sublattice orderings, $\mathbf{M}=(\mathbf{S}_{\rm{A}}, \mathbf{S}_{\rm{B}})=(S^x_{\rm{A}},S^y_{\rm{A}},S^z_{\rm{A}},S^x_{\rm{B}},S^y_{\rm{B}},S^z_{\rm{B}})$.
When taking $J<0$ and $K=0$, an optimal spin configuration is described by a linear combination of the different bases as 
\begin{align}
\mathbf{M}=m^{{\rm{FM}}_x}\mathbf{M}^{{\rm{FM}}_x}+m^{{\rm{AFM}}_y}\mathbf{M}^{{\rm{AFM}}_y}+m^{{\rm{FM}}_z}\mathbf{M}^{{\rm{FM}}_z}, 
\end{align}
 where
\begin{align}
&\mathbf{M}^{{\rm{FM}}_x}=(1,0,0,1,0,0),\\
&\mathbf{M}^{{\rm{AFM}}_y}=(0,1,0,0,-1,0),\\
&\mathbf{M}^{{\rm{FM}}_z}=(0,0,1,0,0,1),
\end{align}
and $\{m^{{\rm{FM}}_x}, m^{{\rm{AFM}}_y}, m^{{\rm{FM}}_z}\}$ are numerical coefficients.
The ground-state energy of the model in Eq.~\eqref{eq1} is minimized as $(m^{{\rm{FM}}_x}, m^{{\rm{AFM}}_y}, m^{{\rm{FM}}_z})=(\sin{\theta},0,\cos{\theta})$ in the case of $D=0$; the ferromagnetic ordering with the moments along the magnetic field direction becomes the ground state.
The introduction of $D$ cants the ferromagnetic moments so as to have the antiferromagnetic moments along the $y$ direction, i.e., $m^{{\rm{AFM}}_y} \neq 0$.
When setting the relative angle of $\mathbf{S}_{\rm{A}}$ and $\mathbf{S}_{\rm{B}}$ as $\phi$, $\phi$ is given by 
\begin{align}
\phi=\arctan\left[\frac{(1-\gamma')D}{(1+\gamma)J}\right],
\end{align}
at zero field.
For instance, the angle $\phi$ is about $3.8$ degrees for $(D,J,K,\gamma,\gamma', H)=(0.2,1,0,0.5,0.5,0)$. 
Meanwhile, $\phi$ becomes smaller while introducing $H$, e.g., $\phi\sim1^{\circ}$ for $H=2.6$, which means that the effect of the canted moments is negligible for large $H$.
In this way, we approximately assume the ferromagnetic state by considering large $H=4$ in Sec.~\ref{ferro1} in the main text.

Next, we investigate an optimal spin configuration for $J>0$ and $H=0$.
When taking into account $D$ and $K$, the ground-state spin configuration is given by
\begin{align}
&\mathbf{M}^{{\rm{C}}}=\cos\left(\frac{\phi}{2}\right)\mathbf{M}^{{\rm{FM}}_x}+\sin\left(\frac{\phi}{2}\right)\mathbf{M}^{{\rm{AFM}}_y},
\end{align}
or
\begin{align}
&\mathbf{M}^{{\rm{AFM}}_z}=(0,0,1,0,0,-1),
\end{align}
depending on $D$ and $K$, except for the trivial degenerated state.
The larger $D$ ($K$) tends to favor $\mathbf{M}^{{\rm{C}}}$ ($\mathbf{M}^{{\rm{AFM}}_z}$).
The critical value at $K_{\rm{c}}$ is approximately given by  $K\gtrsim (1-\gamma')^2D^2/[2(1+\gamma)^{2}J]$.
Thus, we take a large value of $K > K_{\rm c}$ in the collinear antiferromagnetic structure $\mathbf{M}^{{\rm{AFM}}_z}$ in Sec.~\ref{antiferro1} in the main text.

\section{Three-sublattice orderings in the breathing kagome structure}
\label{GS2}

\begin{figure}[h]
\begin{center}
\includegraphics[width=1.00\linewidth]{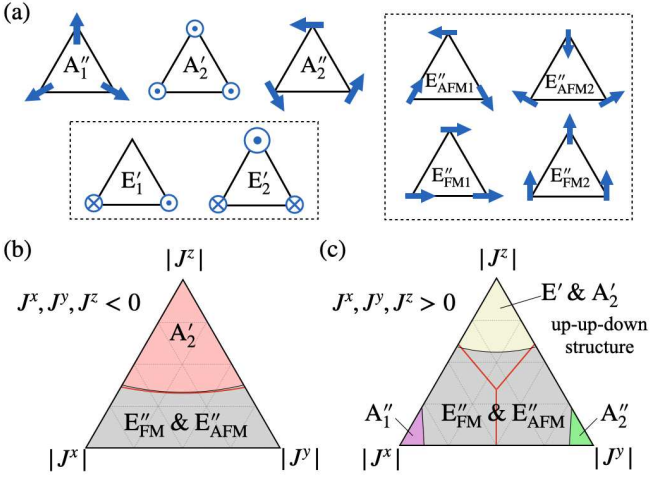}
\end{center}
\caption{(a) Three-sublattice spin structures in each irreducible representation.
(b), (c) The magnetic phase diagrams in (b) the ferromagnetic case ($J^x, J^y, J^z <0$) and (c) the antiferromagnetic case ($J^x, J^y, J^z >0$) where $J^x\equiv J+J^a$ and $J^x\equiv J-J^a$. 
The other parameters are set as $\gamma=\gamma'$, $J^{\perp }=J^{\perp z}$, $J^{\perp a}=0$, $D^{\perp}=0$,  and $D=0.2
$. 
The red curves represent the phase boundaries at $D=0$.
The gray region in (c) disappears at $D=0$.
}
\label{fig8}
\end{figure}

We show the optimal spin configurations within the three-sublattice orderings in the breathing kagome model in Eqs.~\eqref{eq9} and \eqref{eq9-2} in Sec.~\ref{model2}. 
From the group theory, nine independent magnetic spin textures are expressed for the three-sublattice bases as $\mathbf{M}^{\Gamma^{\rm{irrep}}}=(\mathbf{S}^{\Gamma^{\rm{irrep}}}_{\rm{A}},\mathbf{S}^{\Gamma^{\rm{irrep}}}_{\rm{B}},\mathbf{S}^{\Gamma^{\rm{irrep}}}_{\rm{C}})$ in Eqs.~\eqref{a''1}-\eqref{e''2} where each spin configuration is shown in Fig.~\ref{fig8}(a). 
By using the complete orthogonal bases, any three-sublattice spin configurations $\mathbf{\tilde{S}}=(\mathbf{\tilde{S}}_{\rm{A}},\mathbf{\tilde{S}}_{\rm{B}},\mathbf{\tilde{S}}_{\rm{C}})$ are expanded as 
\begin{align}
\label{eq16}
\mathbf{\tilde{S}}=&m^{{\rm{A''_1}}}\mathbf{M}^{{\rm{A''_1}}}+m^{{\rm{A'_2}}}\mathbf{M}^{{\rm{A'_2}}}+m^{{\rm{A''_2}}}\mathbf{M}^{{\rm{A''_2}}}+m^{{\rm{E'_1}}}\mathbf{M}^{{\rm{E'_1}}}\nonumber\\
&+m^{{\rm{E'_2}}}\mathbf{M}^{{\rm{E'_2}}}+m^{\rm{E''_{\rm{FM}1}}}\mathbf{M}^{\rm{E''_{\rm{FM}1}}}+m^{\rm{E''_{\rm{FM}2}}}\mathbf{M}^{\rm{E''_{\rm{FM}2}}}\nonumber\\
&+m^{\rm{E''_{\rm{AFM}1}}}\mathbf{M}^{\rm{E''_{\rm{AFM}1}}}+m^{\rm{E''_{\rm{AFM}2}}}\mathbf{M}^{\rm{E''_{\rm{AFM}2}}}
,\end{align}
where 
$m^{\Gamma^{\rm{irrep}}}$ is obtained by
\begin{align}
m^{\Gamma^{\rm{irrep}}}=\frac{1}{3}\mathbf{\tilde{S}}\cdot \mathbf{M}^{\Gamma^{\rm{irrep}}}
.\end{align}
By substituting $\mathbf{\tilde{S}}_{\eta}$ into Eq.~\eqref{eq15}, the spin Hamiltonian is rewritten as
\begin{align}
\label{eq17}
H=&
N\big[h^{{\rm{A''_1}}}(m^{{\rm{A''_1}}})^2+h^{{\rm{A''_2}}}(m^{{\rm{A''_2}}})^2+h^{{\rm{A'_2}}}(m^{{\rm{A'_2}}})^2\nonumber\\
&+h^{{\rm{E'_1}}}(\mathbf{m}^{E'})^2+h^{\rm{E''_{\rm{AFM}}}}(\mathbf{m}^{\rm{E''_{\rm{AFM}}}})^2+h^{\rm{E''_{\rm{FM}}}}(\mathbf{m}^{\rm{E''_{\rm{FM}}}})^2\nonumber\\
&+2h^{\rm{E''_{\rm{mix}}}} \mathbf{m}^{\rm{E''_{\rm{AFM}}}}\cdot\mathbf{m}^{\rm{E''_{\rm{FM}}}}
\big]
,\end{align}
where $N$ is the number of total unit cells and the coefficients $h^{\Gamma^{\rm{irrep}}}$ are given by
\begin{subequations}
\begin{align}
\label{h1}
&h^{{\rm{A''_1}}}=\frac{3}{4}(1+\gamma)(J^x-3J^y)+\frac{3\sqrt{3}}{2}(1+\gamma')D\nonumber\\
&\ \ \ \ \ \ \ \ +3(J^{\perp}-J^{\perp a}),\\
\label{h2}
&h^{{\rm{A''_2}}}=\frac{3}{4}(1+\gamma)(-3J^x+J^y)+\frac{3\sqrt{3}}{2}(1+\gamma')D\nonumber\\
&\ \ \ \ \ \ \ \ +3(J^{\perp}+J^{\perp a}),\\
\label{h3}
&h^{{\rm{A'_2}}}=3(1+\gamma)J^z+3J^{\perp z},\\
\label{h4}
&h^{{\rm{E'_1}}}=-\frac{3}{2}(1+\gamma)J^z+3J^{\perp z},\\
&h^{\rm{E''_{\rm{AFM}}}}=-\frac{3}{4}(1+\gamma)(J^x+J^y)-\frac{3\sqrt{3}}{2}(1+\gamma')D +3J^{\perp},\\
&h^{\rm{E''_{\rm{FM}}}}=\frac{3}{2}(1+\gamma)(J^x+J^y)+3J^{\perp},\\
&h^{\rm{E''_{\rm{mix}}}}=\frac{3}{4}(1+\gamma)(J^x-J^y)+3J^{\perp a}
,\end{align}
\end{subequations}
with $J^{x}\equiv J+J^{a}$ and $J^{y}\equiv J-J^{a}$. 
It is noted that $D^{\perp}$ does not contribute to the ground-state energy within the three-sublattice orderings.

Then the spin Hamiltonian is diagonalized except for the magnetic orderings belonging to the irreducible representation ${\rm{E''}}$.
For ${\rm{E''}}$ orderings, we choose their linear combination as
\begin{align}
\label{eq19}
\mathbf{m}^{\rm{E''_{\alpha}}}=&\mathbf{m}^{\rm{E''_{\rm{FM}}}}\cos{\chi}-\mathbf{m}^{\rm{E''_{\rm{AFM}}}}\sin{\chi},\\
\mathbf{m}^{\rm{E''_{\beta}}}=&\mathbf{m}^{\rm{E''_{\rm{FM}}}}\sin{\chi}+\mathbf{m}^{\rm{E''_{\rm{AFM}}}}\cos{\chi}
,\end{align}
where $\chi$ is given by
\begin{align}
\label{eq20}
\chi=\frac{1}{2}\tan^{-1}
\left[\frac{h^{\rm{E''_{\rm{mix}}}}}{h^{\rm{E''_{\rm{FM}}}}-h^{\rm{E''_{\rm{AFM}}}}}
\right]
.\end{align}
The corresponding eigenvalues are obtained by
\begin{align}
\label{h5}
h^{\rm{E''_{\alpha}}}=&h^{\rm{E''_{\rm{FM}}}} \cos^2{\chi}+h^{\rm{E''_{\rm{AFM}}}}\sin^2{\chi}-\frac{1}{2}h^{\rm{E''_{\rm{mix}}}}\sin{2\chi} ,\\
\label{h6}
h^{\rm{E''_{\beta}}}=&h^{\rm{E''_{\rm{FM}}}}\sin^2{\chi}+h^{\rm{E''_{\rm{AFM}}}}\cos^2{\chi}+\frac{1}{2}h^{\rm{E''_{\rm{mix}}}}\sin{2\chi}
.\end{align}
By comparing $h^{\Gamma^{\rm{irrep}}}$ in Eqs.~\eqref{h1}-\eqref{h4}, \eqref{h5}, and \eqref{h6}, we can obtain the ground state.
It is noted that we consider up-up-down spin configuration given by
\begin{align}
\label{eq221}
\mathbf{M}^{++-}&=\frac{1}{3}\mathbf{M}^{{\rm{A'_2}}}+\frac{2\sqrt{2}}{3}\mathbf{M}^{{\rm{E'_2}}},
\end{align}
instead of the $\rm{E'}$ spin orderings so as to satisfy $|\mathbf{S}_{i\eta}|=1$. 
We also consider the energetically degenerate up-down-up spin configuration when investigating magnon dispersions:
\begin{align}
\label{eq222}
\mathbf{M}^{+-+}&=\frac{1}{3}\mathbf{M}^{{\rm{A'_2}}}+\sqrt{\frac{2}{3}}\mathbf{M}^{{\rm{E'_1}}}+\frac{\sqrt{2}}{3}\mathbf{M}^{{\rm{E'_2}}}
.
\end{align}
Figures~\ref{fig8}(b) and \ref{fig8}(c) represent the obtained phase diagram in the cases of the ferromagnetic ($J^x, J^y, J^z<0$) and antiferromagnetic ($J^x, J^y, J^z>0$) interactions, respectively.
The inter-plane interactions are taken at $J^{\perp }=J^{\perp z}$, $J^{\perp a}=0$, $D^{\perp}=0$.
The similar variational analysis on a triangular cluster has also been presented in Refs.~\cite{PhysRevB.96.205126, PhysRevB.103.174425}.
For simplicity, we neglect a possibility of a classical spin liquid state, which have been found in the ground state for the isotropic spin model in a two-dimensional kagome structure~\cite{kano1953antiferromagnetism, PhysRevLett.59.1629, PhysRevB.45.7536}.

The phase boundaries depend on the inter-plane interactions.
The larger $J^{\perp z}$ tends to favor the $\rm{A}'_2$ ferromagnetic state in Fig.~\ref{fig8}(b) and the up-up-down state in Fig.~\ref{fig8}(c).
 The interactions of $J^{\perp x} (\equiv J^{\perp}+J^{\perp a})$ and $J^{\perp y} (\equiv J^{\perp}-J^{\perp a})$ tend to extend the region of the $\rm{E}''$ state in Fig.~\ref{fig8}(b), and the $\rm{A}''_2$ and  $\rm{A}''_1$ antiferromagnetic states in Fig.~\ref{fig8}(c) respectively.

In the main text, we use the parameters as $(J, J^z, J^a, D, J^{\perp}, J^{\perp z }, J^{\perp a})=(-0.8, -1, 0, 0.2, 0, 0, 0)$ for the ferromagnetic state with the moments along the $z$ axis in Sec.~\ref{a2}, as $(J, J^z, J^a, D, J^{\perp}, J^{ \perp z}, J^{\perp a})=(0.8, 1, 0, 0.2, 0, -1, 0)$ for the up-up-down state in Sec.~\ref{e}, as $(J, J^z, J^a, D, J^{\perp}, J^{ \perp z}, J^{\perp a})=(1, 0.8, -0.5, 0, -1, -0.8, -0.5)$ for the noncollinear antiferromagnetic state in Sec.~\ref{a22}, as $(J, J^z, J^a, D, J^{\perp}, J^{\perp z}, J^{\perp a})=(-1, -0.5, 0, 0, -1, -0.5, 0)$ for the ferromagnetic state with the moments along the $x$ axis in Sec.~\ref{e2}.

\section{Bogoliubov Hamiltonian in the breathing kagome structure}
\label{SW2}
We show the Bogoliubov Hamiltonian obtained from the model $H^{\rm intra-plane}+H^{\rm inter-plane}$ in Eqs.~\eqref{eq9} and \eqref{eq9-2} by supposing the three-sublattice ordering $\mathbf{S}_{\eta}=S(\sin{\theta_{\eta}}\cos{\phi_{\eta}},\sin{\theta_{\eta}}\sin{\phi_{\eta}}, \cos{\theta_{\eta}})$.
We use the rotated spin frame so that the local quantization axis is along the $z$ direction, which is achieved by
\begin{align}
\label{eq2_2}
\begin{pmatrix}
S^{x}_{i\eta}\\
S^{y}_{i\eta}\\
S^{z}_{i\eta}
\end{pmatrix}
=
R_z(\phi_{\eta})R_y (\theta_{\eta})
\begin{pmatrix}
\tilde{S}^{x}_{i\eta}
\\
\tilde{S}^{y}_{i\eta}
\\
\tilde{S}^{z}_{i\eta}
\end{pmatrix},
\end{align}
where $R_z(\phi)$ and $R_y(\theta)$ are the rotation matrices around the $z$ and $y$ axes, respectively.
By introducing the boson operators $a_{i}$, $b_{i}$, and $c_{i}$ for the A, B, and C sublattices, and performing the Fourier transformation, we derive the Bogoliubov Hamiltonian in $\mathbf{q}$ space, which is given by
\begin{align}
\label{eq23}
H=\frac{S}{2}
\sum_{\mathbf{q}}
(\xi^{\dagger}_{\mathbf{q}}, \xi_{-\mathbf{q}})
\begin{pmatrix}
X_{\mathbf{q}}&Y_{\mathbf{q}}\\
Y^{\dagger}_{\mathbf{q}}&X^{*}_{-\mathbf{q}}
\end{pmatrix}
\begin{pmatrix}
\xi_{\mathbf{q}}\\
\xi^{\dagger}_{-\mathbf{q}}
\end{pmatrix}
,\end{align}
where $\xi^{\dagger}_{\mathbf{q}}\equiv(a^{\dagger}_{\mathbf{q}},b^{\dagger}_{\mathbf{q}}, c^{\dagger}_{\mathbf{q}})$.
$X_{\mathbf{q}}$ and $Y_{\mathbf{q}}$ are the $3\times 3$ matrices, which are given by
\begin{align}
\label{eq24}
X_{\mathbf{q}}=&
\begin{pmatrix}
Z_{\rm{A}{\mathbf{q}}}&F_{\rm{AB}{\mathbf{q}}}&F^{*}_{\rm{CA}{\mathbf{q}}}\\
F^{*}_{\rm{AB}{\mathbf{q}}}&Z_{\rm{B}{\mathbf{q}}}&F_{\rm{BC}{\mathbf{q}}}\\
F_{\rm{CA}{\mathbf{q}}}&F^{*}_{\rm{BC}{\mathbf{q}}}&Z_{\rm{C}{\mathbf{q}}}
\end{pmatrix},\\
Y_{\mathbf{q}}=&
\begin{pmatrix}
Z'_{\rm{A}{\mathbf{q}}}&G_{\rm{AB}{\mathbf{q}}}&G_{\rm{CA}{-\mathbf{q}}}\\
G_{\rm{AB}-{\mathbf{q}}}&Z'_{\rm{B}{\mathbf{q}}}&G_{\rm{BC}{\mathbf{q}}}\\
G_{\rm{CA}{\mathbf{q}}}&G_{\rm{BC}{-\mathbf{q}}}&Z'_{\rm{C}{\mathbf{q}}}
\end{pmatrix}
,\end{align}
with
\begin{align}
\label{eq25}
F_{\eta \eta'{\mathbf{q}}}
=&
\left(\tilde{J}_{{\rm{\eta\eta'}}}-i\tilde{D}_{{\rm{\eta\eta'}}}\right)
e^{+}_{{\rm{\eta}} {\rm{\eta'}}}
+\left(\tilde{J'}_{{\rm{\eta\eta'}}}-i\tilde{D'}_{{\rm{\eta\eta'}}}\right)
e^{-}_{{\rm{\eta}} {\rm{\eta'}}},\\
G_{\eta \eta'{\mathbf{q}}}
=&
\left(\tilde{J}_{{\rm{\eta\eta'}}}^{a}+i\tilde{K}_{{\rm{\eta\eta'}}}\right)e^{+}_{{\rm{\eta}} {\rm{\eta'}}}
+\left(\tilde{J'}_{{\rm{\eta\eta'}}}^{a}+i\tilde{K'}_{{\rm{\eta\eta'}}}\right)e^{-}_{{\rm{\eta}} {\rm{\eta'}}},\\
Z_{{\rm A}{\mathbf{q}}}=&-\tilde{J}^{z}_{\rm{AB}}-\tilde{J'}^{z}_{\rm{AB}}-\tilde{J}^{z}_{\rm{CA}}-\tilde{J'}^{z}_{\rm{CA}}-2\tilde{J}^{\perp z}_{\rm{AA}}\nonumber\\
&+2\tilde{J}_{{\rm{AA}}}^{\perp }\cos(\mathbf{q} \cdot \boldsymbol{\rho}_{{\rm{A}} {\rm{A}}})
+2\tilde{D}^{\perp}_{{\rm{AA}}}\sin(\mathbf{q} \cdot \boldsymbol{\rho}_{{\rm{A}} {\rm{A}}}),\\
Z_{{\rm B}{\mathbf{q}}}=&-\tilde{J}^{z}_{\rm{AB}}-\tilde{J'}^{z}_{\rm{AB}}-\tilde{J}^{z}_{\rm{BC}}-\tilde{J'}^{z}_{\rm{BC}}-2\tilde{J}^{\perp z}_{\rm{BB}}\nonumber\\
&+2\tilde{J}_{{\rm{BB}}}^{\perp}\cos(\mathbf{q} \cdot \boldsymbol{\rho}_{{\rm{B}} {\rm{B}}})
+2\tilde{D}^{\perp}_{{\rm{BB}}}\sin(\mathbf{q} \cdot \boldsymbol{\rho}_{{\rm{B}} {\rm{B}}}),\\
Z_{{\rm C}{\mathbf{q}}}=&-\tilde{J}^{z}_{\rm{BC}}-\tilde{J'}^{z}_{\rm{BC}}-\tilde{J}^{z}_{\rm{CA}}-\tilde{J'}^{z}_{\rm{CA}}-2\tilde{J}^{\perp z}_{\rm{CC}}\nonumber\\
&+2\tilde{J}_{{\rm{CC}}}^{\perp }\cos(\mathbf{q} \cdot \boldsymbol{\rho}_{{\rm{C}} {\rm{C}}})
+2\tilde{D}^{\perp}_{{\rm{CC}}}\sin(\mathbf{q} \cdot \boldsymbol{\rho}_{{\rm{C}} {\rm{C}}}),\\
Z'_{\eta{\mathbf{q}}}=&\tilde{J}_{{\rm{\eta\eta}}}^{\perp a}\cos(\mathbf{q} \cdot \boldsymbol{\rho}_{{\rm{\eta}} {\rm{\eta}}}),
\end{align}
where $e^{\pm}_{{\rm{\eta}} {\rm{\eta'}}}=e^{\pm i \mathbf{q} \cdot \boldsymbol{\rho}_{{\rm{\eta}} {\rm{\eta'}}}}$ and $\boldsymbol{\rho}_{{\rm{\eta}} {\rm{\eta'}}}$ are the bond vectors from sublattice $\eta$ to $\eta'$ defined in Eqs.~\eqref{bond1}-\eqref{bond2}. 
The interactions $\tilde{J}_{{\rm{\eta\eta'}}}$, $\tilde{J}_{{\rm{\eta\eta'}}}^{a}$, $\tilde{J}_{{\rm{\eta\eta}}}^{z}$, $\tilde{K}_{{\rm{\eta\eta'}}}$ and $\tilde{D}_{{\rm{\eta\eta'}}}$ represent the renormalized interactions within the upward triangle, which are given by
\begin{align}
\label{eq261}
\tilde{J}_{\eta \eta'}
=&\frac{\cos{\theta_{\eta }}\cos{\theta_{\eta'}}+1}{2}[J \cos{\phi^{-}_{\eta \eta'}}-D\sin{\phi^{-}_{\eta \eta'}}]\nonumber\\
&+\frac{\cos{\theta_{\eta }}\cos{\theta_{\eta'}}-1}{2}J^a \cos{(\phi^{+}_{\eta \eta'}+\chi_{\eta \eta'})}\nonumber\\
&+\frac{J^z}{2}\sin{\theta_{\eta }}\sin{\theta_{\eta'}},\\
\tilde{J}_{\eta \eta'}^{a}
=&\frac{\cos{\theta_{\eta }}\cos{\theta_{\eta'}}-1}{2}[J \cos{\phi^{-}_{\eta \eta'}}-D\sin{\phi^{-}_{\eta \eta'}}]\nonumber\\
&+\frac{\cos{\theta_{\eta }}\cos{\theta_{\eta'}}+1}{2}J^a \cos{(\phi^{+}_{\eta \eta'}+\chi_{\eta \eta'})}\nonumber\\
&+\frac{J^z}{2}\sin{\theta_{\eta }}\sin{\theta_{\eta'}},\\
\tilde{J}_{\eta \eta'}^{z}
=&
\sin{\theta_{\eta }}\sin{\theta_{\eta'}}[J\cos{\phi^{-}_{\eta \eta'}}-D\sin{\phi^{-}_{\eta \eta'}}\nonumber\\
&+J^a \cos{(\phi^{+}_{\eta \eta'}+\chi_{\eta \eta'})}]+J^z\cos{\theta_{\eta }}\cos{\theta_{\eta'}},\\
\tilde{K}_{\eta \eta'}
=&
\frac{\cos{\theta_{\eta}}-\cos{\theta_{\eta'}}}{2}\left[J\sin{\phi^{-}_{\eta \eta'}}+D\cos{\phi^{-}_{\eta \eta'}}\right]\nonumber\\
&-\frac{\cos{\theta_{\eta}}+\cos{\theta_{\eta'}}}{2} J^a\sin{(\phi^{+}_{\eta \eta'}+\chi_{\eta \eta'})},\\
\label{eq262}
\tilde{D}_{\eta \eta'}
=&
\frac{\cos{\theta_{\eta}}+\cos{\theta_{\eta'}}}{2}\left[ J\sin{\phi^{-}_{\eta \eta'}}+D\cos{\phi^{-}_{\eta \eta'}}\right]\nonumber\\
&-\frac{\cos{\theta_{\eta}}-\cos{\theta_{\eta'}}}{2} J^a \sin{(\phi^{+}_{\eta \eta'}+\chi_{\eta \eta'})}
,\end{align}
where $\phi^{\pm}_{\eta \eta'}=\phi_{\eta}\pm\phi_{\eta'}$, $\chi_{\rm{AB}}=0$, $\chi_{\rm{BC}}=2\pi/3$, and $\chi_{\rm{CA}}=4\pi/3$.
$\tilde{J}'_{{\rm{\eta\eta'}}}$, $\tilde{J'}_{{\rm{\eta\eta'}}}^{a}$, $\tilde{J'}_{{\rm{\eta\eta'}}}^{z}$, $\tilde{K}'_{{\rm{\eta\eta'}}}$ and $\tilde{D'}_{{\rm{\eta\eta'}}}$ are the interactions within the downward triangle, which are obtained by reading $(J,J^a,J^z,D)$ in Eqs.~\eqref{eq261}-\eqref{eq262} with $(\gamma J, \gamma J^a, \gamma J^z, \gamma' D)$.
$\tilde{J}^{\perp}_{{\rm{\eta\eta}}}$, $\tilde{J}^{\perp a}_{{\rm{\eta\eta}}}$, $\tilde{J}^{\perp z}_{{\rm{\eta\eta}}}$, $\tilde{K}^{\perp}_{{\rm{\eta\eta}}}$ and $\tilde{D}^{\perp}_{{\rm{\eta\eta}}}$ represent the interactions between the breathing kagome planes, which are given by
\begin{align}
\label{27}
\tilde{J}^{\perp }_{\eta \eta}
=&
\frac{\cos^2{\theta_{\eta }}+1}{2}J^{\perp}+\frac{\cos^2{\theta_{\eta }}-1}{2}J^{\perp a}\cos(2\phi_\eta+\chi_{\eta})\nonumber\\
&+J^{\perp z}\sin^2{\theta_{\eta }},\\
\tilde{J}^{\perp a}_{\eta \eta}
=&
\frac{\cos^2{\theta_{\eta }}-1}{2}J^{\perp}+\frac{\cos^2{\theta_{\eta }}+1}{2}J^{\perp a}\cos(2\phi_\eta+\chi_{\eta})\nonumber\\
&+J^{\perp z}\sin^2{\theta_{\eta }},\\
\tilde{J}^{\perp z}_{\eta \eta}
=&
\sin^2{\theta_{\eta }}[J^{\perp}+J^{\perp a}\cos(2\phi_\eta+\chi_{\eta})]
+J^{\perp z}\cos^2{\theta_{\eta }},\\
\tilde{K}^{\perp}_{\eta \eta}
=&
0,\\
\tilde{D}^{\perp}_{\eta \eta}
=&
D^{\perp}\sin{\theta_{\eta}}\cos(\phi_{\eta}-\chi^{\perp}_{\eta})
,\end{align}
where $\chi^{\perp}_{\rm{A}}=2\pi/3$, $\chi^{\perp}_{\rm{B}}=4\pi/3$, and $\chi^{\perp}_{\rm{C}}=0$.
The magnon dispersions in Secs.~\ref{a2}, \ref{e}, \ref{a22}, and \ref{e2} are obtained by substituting the corresponding spin structures into Eq.~(\ref{eq23}) and performing the numerical Bogoliubov transformation.

\section{Nonreciprocal magnons for the other two ordered states}
\label{other}

In this section, we show the nonreciprocal magnons for the other spin configurations, which are not mentioned in the main text. 
We consider the two cases: The one is the up-down-up state in Appendix~\ref{updownup} and the other is the ferromagnetic state with the moments along the $y$ direction in Appendix~\ref{ferrox}.

\begin{figure}[h]
\begin{center}
\includegraphics[width=0.9\linewidth]{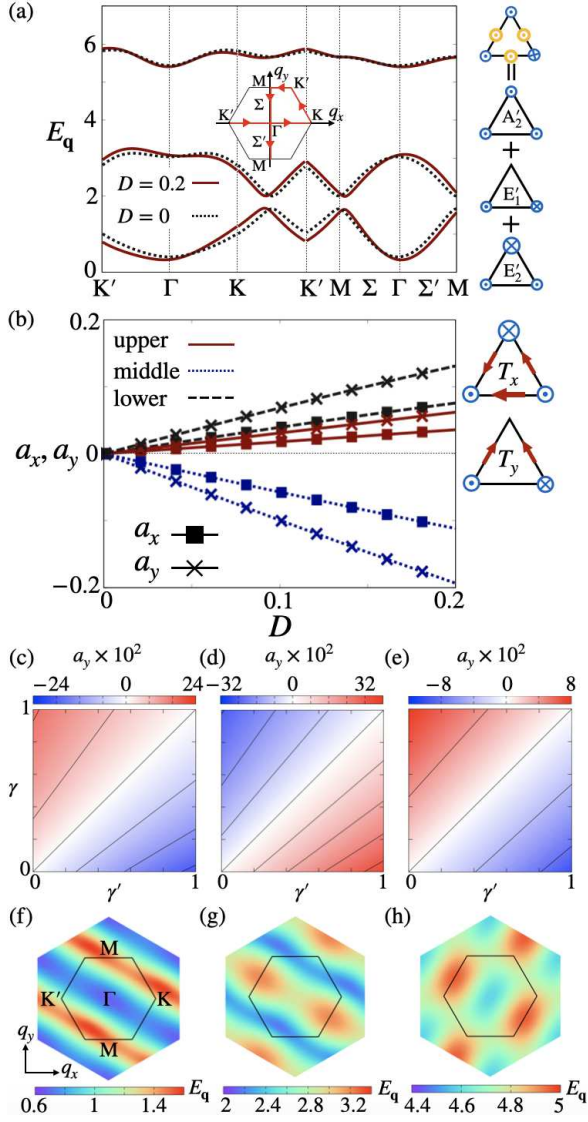}
\end{center}
\caption{
(a) Magnon dispersions in the $\mathbf{M}^{+-+}$ state along the high symmetrical line (see the inset).
The spin configuration is shown in the right panel.
The results are calculated at $(J, J^z, J^{\perp z}, \gamma, \gamma' )=(0.8, 1, -1, 1, 0.2 )$ for $D=0$ and $D=0.2$ (other parameters are zero).
(b) $D$ dependence of $a_{x}$ (square symbol) and $a_{y}$  (cross symbol) for the three branches in (a).
The schematic pictures of the CMT dipoles $T_{x}$ and $T_{y}$ corresponding to $a_{x}$ and $a_{y}$ are shown in the right panel.
(c)-(e) Intensity plot of $a_{y}$ while varying $\gamma$ and $\gamma'$ for (c) lower, (d) middle, and (e) upper branches. 
Contour lines are drawn every $0.08$ for (c) and (d), and $0.04$ for (e).
(f)-(h) The color plot of magnon dispersions for (f) lower, (g) middle, and (h) upper  branches calculated at $\gamma=\gamma'=0.5$.
}
\label{fig4-2}
\end{figure}

\begin{figure}[h]
\begin{center}
\includegraphics[width=0.9\linewidth]{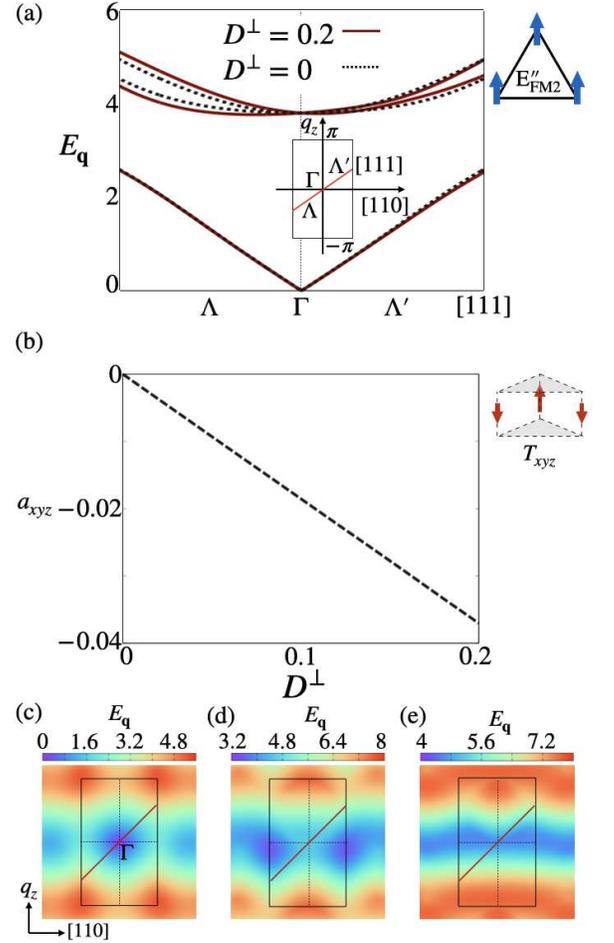}
\end{center}
\caption{
(a) Magnon dispersions in the $\mathbf{M}^{\rm{E''_{FM2}}}$ 
state along the [111] line (see the inset).  
The spin configuration is shown in the right panel.
The results for $D^{\perp}=0$ and $D^{\perp}=0.2$ are shown at $(J, J^z, J^{\perp}, J^{\perp z}, \gamma)=(-1, -0.5, -1, -0.5, 0.5)$ (other parameters are zero).
(b) $D^{\perp}$ dependence of $a_{xyz}$ for the lower branch in (a).
The schematic picture of the CMT dipole $T_{xyz}$ corresponding to nonzero $a_{xyz}$ is shown in the right panel.
(c)-(e) The color plot of magnon dispersions for (c) lower, (d) middle, and (e) upper  branches.
}
\label{fig6-1}
\end{figure}

\subsection{Up-down-up state}
\label{updownup}

We consider the up-down-up structure $\mathbf{M}^{+-+}$ given in Eq.~\eqref{eq222}, which is connected with the $\mathbf{M}^{++-}$ state by the threefold rotational operation.
The $\mathbf{M}^{+-+}$ state is constructed by the superpositions of the bases belonging to $\rm{A'_{2}}$, $\rm{E'}_1$, and $\rm{E'}_2$, as shown in the right panel in Fig.~\ref{fig4-2}(a).
This state is stabilized for the same parameters as $\mathbf{M}^{++-}$ in Sec.~\ref{e} as a metastable state.

Figure~\ref{fig4-2} shows the magnon dispersions in the $\mathbf{M}^{+-+}$ state at $(J, J^z, D, J^{\perp z}, \gamma, \gamma' )=(0.8, 1, 0.2,  -1, 1, 0.2 )$.
The dispersions show the bottom shift along both the $\rm{K'}$-$\rm{\Gamma}$-$\rm{K}$ line and the $\rm{M}$-$\rm{\Gamma}$ line as shown in Fig.~\ref{fig4-2}(a), which indicates the formation of the CMT dipoles $T_x$ and $T_y$ with nonzero $a_{x}$ and $a_{y}$ [Fig.~\ref{fig4-2}(b)].
The ratio of $a_y$ and $a_x$ is given by $\arctan(a_{y}/a_{x})=-2\pi/3$, which is understood from the fact that the $\mathbf{M}^{+-+}$ state is obtained by the threefold rotational operation to the $\mathbf{M}^{++-}$ state. 
We plot the breathing parameter dependence of $a_{y}$ in Eq.~\eqref{eq13} in Figs.~\ref{fig4-2}(c)-\ref{fig4-2}(e) and the magnon dispersions in Figs.~\ref{fig4-2}(f)-\ref{fig4-2}(h).

\subsection{Ferromagnetic state with the moments along the $y$ axis}
 \label{ferrox}

We show the magnon band dispersions in the ferromagnetic state with the moments along the $y$ direction, which belongs to the $\rm{E''}$ irreducible representation.
The state is stabilized in the same parameter region as the ferromagnetic state with the moments along the $x$ direction in Sec.~\ref{e2}.

The magnon dispersions at $(J, J^z, J^{\perp}, J^{\perp z}, D^{\perp}  \gamma)=(-1, -0.5, -1, -0.5, 0.5, 0.5 )$ show the asymmetric band inclination in the [111] direction as shown in Fig.~\ref{fig6-1}(a), which is understood from the emergence of the CMT octupole $T_{xyz}$.
Figure~\ref{fig6-1}(b) shows that $a_{xyz}$ behaves linearly while increasing $D^{\perp}$. 
We plot the magnon dispersions in Figs.~\ref{fig6-1}(c)-\ref{fig6-1}(e).

\bibliography{refkagome}

\end{document}